\documentclass[aip,jcp,amsmath,amssymb,floatfix,reprint,citeautoscript]{revtex4-2}

\usepackage[utf8]{inputenc}
\usepackage{graphicx}
\usepackage{wrapfig}
\usepackage{placeins}
\usepackage{bibunits}
\usepackage{siunitx}
\usepackage{xr-hyper}
\usepackage[colorlinks,allcolors=black,citecolor=blue,urlcolor=blue]{hyperref}
\usepackage[mathlines]{lineno}

\graphicspath{{figures/}}
\hypersetup{colorlinks=true, urlcolor=blue}

\newcommand{\vect}[1]{\mathbf{#1}}

\defaultbibliography{paper,revtex-control}
\defaultbibliographystyle{aipnum4-2}

\makeatletter
\def\@email#1#2{%
 \endgroup
 \patchcmd{\titleblock@produce}
  {\frontmatter@RRAPformat}
  {\frontmatter@RRAPformat{\produce@RRAP{*#1\href{mailto:#2}{#2}}}\frontmatter@RRAPformat}
  {}{}
}%
\makeatother

\begin{document}

\def\mytitle{
Multi-head committees enable direct uncertainty prediction for atomistic foundation models
}
\title{\mytitle}

\author{Hubert Beck}
\affiliation{
Charles University, Faculty of Mathematics and Physics, Ke Karlovu 3, 121 16 Prague 2, Czech Republic
}

\author{Pavol Simko}
\affiliation{
Charles University, Faculty of Mathematics and Physics, Ke Karlovu 3, 121 16 Prague 2, Czech Republic
}

\author{Lars L. Schaaf}
\affiliation{
Cavendish Laboratory, Department of Physics, University of Cambridge, Cambridge CB3 0HE, U.K.
}
\affiliation{
Lennard-Jones Centre, University of Cambridge, Trinity Ln, Cambridge CB2 1TN, U.K.
}

\author{Ondrej Marsalek}
\email{ondrej.marsalek@matfyz.cuni.cz}
\affiliation{
Charles University, Faculty of Mathematics and Physics, Ke Karlovu 3, 121 16 Prague 2, Czech Republic
}

\author{Christoph Schran}
\email{cs2121@cam.ac.uk}
\affiliation{
Cavendish Laboratory, Department of Physics, University of Cambridge, Cambridge CB3 0HE, U.K.
}
\affiliation{
Lennard-Jones Centre, University of Cambridge, Trinity Ln, Cambridge CB2 1TN, U.K.
}

\date{\today}

\begin{abstract}

Machine learning potentials have become a standard tool for atomistic materials modeling.
While models continue to become more generalizable, an open challenge relates to efficient uncertainty predictions for active learning and robust error analysis.
In this work, we utilize MACE and its multi-head mechanism to implement a committee neural network potential for message-passing architectures, where the committee comprises multiple output modules attached to the same atomic environment descriptors.
As with traditional committees of independent networks, the standard deviation of the predictions functions as an estimate of the model's uncertainty.
We show for a range of datasets in custom-build models that the uncertainty of the force predictions correlates well with the true errors.
We subsequently apply this concept to foundation models, specifically MACE-MP-0, where we train only the newly attached output heads while keeping the remaining part of the model fixed.
We use this approach in an active learning workflow to condense the training set of the foundation model to just 5\% of its original size.
The foundation model multi-head committee trained on the condensed training set enables reliable uncertainty estimation without any substantial decrease in prediction accuracy.

\end{abstract}

{\maketitle}

\begin{bibunit}

\nocite{revtex-control}

\section{Introduction}
Over the past decade, the field of machine learning potentials (MLPs) has made multiple significant advances~\cite{Behler2021/10.1021/acs.chemrev.0c00868, MartinBarrios2024/10.1002/qua.27389}.
Architectures with fixed atomic environment descriptors (AEDs), such as Gaussian approximation potentials (GAP)~\cite{Bartok2010/10.1103/PhysRevLett.104.136403} and Behler--Parrinello high-dimensional neural network potentials~\cite{Behler2007/10.1103/PhysRevLett.98.146401}, established MLPs as a powerful alternative to traditional means of obtaining the potential energy surface (PES) for molecular dynamics (MD) simulations, such as empirical force fields and ab initio methods~\cite{Behler2016/10.1063/1.4966192, Gastegger2017/10.1039/C7SC02267K, Unke2021/10.1021/acs.chemrev.0c01111, Mortazavi2023/10.1039/D3MH00125C}.
The next step in the evolution was graph neural network potentials, where the fixed AEDs were replaced by a message-passing graph neural network, which made the representation of the local atomic environment a learnable feature of the model.
While SchNet~\cite{Schutt2017/AiNIP30} was the first notable network of this kind, the introduction of many-body terms~\cite{Drautz2019/10.1103/PhysRevB.99.014104}, higher-order tensor features, and equivariant kernels~\cite{Kondor2018/10.48550/arXiv.1802.03690, Thomas2018/10.48550/arXiv.1802.08219, Geiger2022/10.48550/arXiv.2207.09453} in packages such as NequIP~\cite{Batzner2022/10.1038/s41467-022-29939-5}, Allegro~\cite{Musaelian2023/10.1038/s41467-023-36329-y}, MACE~\cite{Batatia2022/MACE}, TeaNet~\cite{Takamoto2022/10.1016/j.commatsci.2022.111280}, or AlphaNet~\cite{Yin2025/10.48550/arXiv.2501.07155} brought them to the forefront of recent attention.
These MLPs have raised the standard, both in terms of prediction accuracy and training data efficiency~\cite{Leimeroth2025/10.48550/arXiv.2505.02503}.
Furthermore, they are capable of handling extensive and diverse training data, covering a wide range of different elements and systems~\cite{Chen2022/10.1038/s43588-022-00349-3, Deng2023/10.1038/s42256-023-00716-3, Batatia2023/10.48550/arXiv.2401.00096}.

These versatile characteristics of modern MLP architectures have led to the recent development of foundation models.
These are trained on very large datasets that span chemical compound space across the periodic table and are capable of running stable MD out of the box for a broad spectrum of systems, even those that might not be covered directly by the training set.
This is in contrast to the more common MLPs custom-made for a class of related systems, which are trained specifically on structures in the same domain as those encountered at inference time.
In a combined approach, foundation models can be fine-tuned on specific molecules or materials of interest to increase their accuracy~\cite{Allen2024/10.1038/s41524-024-01339-x, Kaur2025/10.1039/D4FD00107A}.
New foundation models such as M3GNet~\cite{Chen2022/10.1038/s43588-022-00349-3}, CHGNet~\cite{Deng2023/10.1038/s42256-023-00716-3}, MACE-MP-0~\cite{Batatia2023/10.48550/arXiv.2401.00096}, GNoME~\cite{Merchant2023/10.1038/s41586-023-06735-9}, MatterSim~\cite{Yang2024/10.48550/arXiv.2405.04967}, grACE-2L~\cite{Bochkarev2024/10.1103/PhysRevX.14.021036}, SevenNet-MF-ompa~\cite{Kim2024/10.1021/jacs.4c14455}, and eSEN-30M~\cite{Fu2025/10.48550/arXiv.2502.12147} are getting released by both academic and for-profit entities on an almost weekly basis, which further demonstrates the impact of this development.

While proving to be a major leap in the field for the exploration of new material chemistry and physics, foundation models are in most applications only qualitatively correct and can still show unphysical behavior~\cite{Focassio2025/10.1021/acsami.4c03815}.
In this context, it would be advantageous to have easy ways of quantifying a foundation model's uncertainty, which could be used to assess and monitor the accuracy of its predictions and identify failures more directly~\cite{Imbalzano2021/10.1063/5.0036522, Dai2025/10.1515/revce-2024-0028}.
A range of methods and workflows to address this issue is known in the context of MLPs~\cite{Gawlikowski2023/10.1007/s10462-023-10562-9}, but they have not yet seen widespread adoption in foundation models, given the extensive cost of training.
Established methods for uncertainty prediction in MLPs include a last-layer approximation of prediction rigidities~\cite{Bigi2024/10.1088/2632-2153/ad805f}, the prediction of confidence intervals using quantile regression~\cite{Bilbrey2025/10.1038/s41524-025-01572-y}, Gaussian mixture models trained on atomic environment descriptors~\cite{Zhu2023/10.1063/5.0136574}, a model-free estimator based on information entropy~\cite{Schwalbe-Koda2025/10.1038/s41467-025-59232-0}, committee neural network~\cite{Schran2020/10.1063/5.0016004, Carrete2023/10.1063/5.0146905} potentials, and shallow ensembles~\cite{Kellner2024/10.1088/2632-2153/ad594a}.
Once available, uncertainties can be used for active learning, a data-driven workflow to find the most relevant training structures out of a large set of candidates~\cite{Schran2020/10.1063/5.0016004, Carrete2023/10.1063/5.0146905, Schaaf2023/10.1038/s41524-023-01124-2, Holzmuller2023/activeLearning}.
Such uncertainty estimates can then be used to monitor the reliability of a model's prediction or, in the context of active learning, to condense and optimize training sets and to reduce the number of
necessary reference electronic structure calculations.
Having this uncertainty for foundation models would be particularly valuable.
Considering the enormous effort required to train a foundation model, a solution to the uncertainty quantification problem should take advantage of the capacity of the foundation model while leaving its predictive power untouched.
Furthermore, it should not add a substantial computational cost to the model and allow for a calculation of the uncertainty on the fly.

A suitable framework for implementing an uncertainty measure guided by these considerations is MACE~\cite{Batatia2022/MACE}, a leading implementation of multi-ACE~\cite{Batatia2025/10.1038/s42256-024-00956-x}, due to its well-established foundation models and fine-tuning workflow.
MACE, which combines the atomic cluster expansion~\cite{Drautz2019/10.1103/PhysRevB.99.014104} with message-passing graph neural network potentials, was initially designed for MLPs trained from scratch for specific systems.
Eventually, it became one of the first packages to embrace the concept of foundation models in atomistic simulations~\cite{Batatia2023/10.48550/arXiv.2401.00096}.
Today, numerous variations and generations of MACE foundation models based on different datasets are available.
Recently, MACE has been extended with a multi-head architecture~\cite{Batatia2023/10.48550/arXiv.2401.00096} that allows for multiple output modules to be attached to a shared block of message-passing layers, which form the AEDs.
This enables, for example, efficient training of a single model to multiple different reference methods, as the AEDs are trained on all training structures, and the output heads only on the structures corresponding to a certain electronic structure method.
The most common usage of the mechanism is the fine-tuning of foundation models.

Here, we adapt the MACE multi-head framework to enable uncertainty quantification by building committee models with a shared description of the local atomic environment and individual output heads forming the committee members.
First, we demonstrate that the force-disagreement of a multi-head committee (MHC) serves as a good quantification of the model uncertainty using established datasets spanning from gas-phase molecules to condensed-phase liquids in custom-made MLPs.
Using these simple, focused datasets, we investigate different strategies of distributing the training data between the output heads, examine the impact the committee has on the prediction accuracy, and compare the MHC with a naive committee of independent MACE models.
We then adapt the MHC approach to equip foundation models with a direct uncertainty prediction.
Namely, we use the uncertainty measure in a query by committee (QbC) active learning workflow to condense the large MPtrj training dataset of the MACE-MP-0b foundation model.
We train new output heads on the condensed dataset to form an MHC, which yields both a prediction and an uncertainty estimate.
We show that this uncertainty estimate correlates well with the actual error.
When comparing this MHC with the original foundation model, we observe only a marginal degradation of prediction accuracy.
Testing other strategies to condense the training data shows similar, but slightly inferior, results.
This strategy of condensing the training data without significantly compromising the foundation model's predictive power in neither precision nor generality indicates a future pathway to upgrade the reference method of the models to higher rungs on Jacob's ladder, such as hybrid DFT or even beyond.

\section{Methods\label{sec:methods}}
In order to enable simple uncertainty prediction within the MACE framework, we have implemented an MHC architecture.
The general idea of such a committee has previously been outlined by Kellner et al.~\cite{Kellner2024/10.1088/2632-2153/ad594a}, and was explored for MACE foundation models with an emphasis on energies by Bilbrey et al.~\cite{Bilbrey2025/10.1038/s41524-025-01572-y}.
Leveraging the existing multi-head functionality within MACE, initially conceptualized in the context of fine-tuning~\cite{Batatia2023/10.48550/arXiv.2401.00096}, we train multiple output heads to different subsets of the total training set.
We lay out two options for distributing the training data:
One evenly distributes the full training set between the output heads (``disjoint''), while the other picks the subsets randomly and independently of each other from the full training set (``overlapping'').
Having multiple readouts enables us to use the standard deviation between head predictions for uncertainty quantification, similar to committee neural network potentials, while requiring little additional architectural overhead.
In addition, this design choice makes it very easy to extend a packaged foundation model without requiring retraining.

\subsection{Multi-head committee}

\begin{figure}
\centering
\includegraphics[width=0.95\columnwidth]{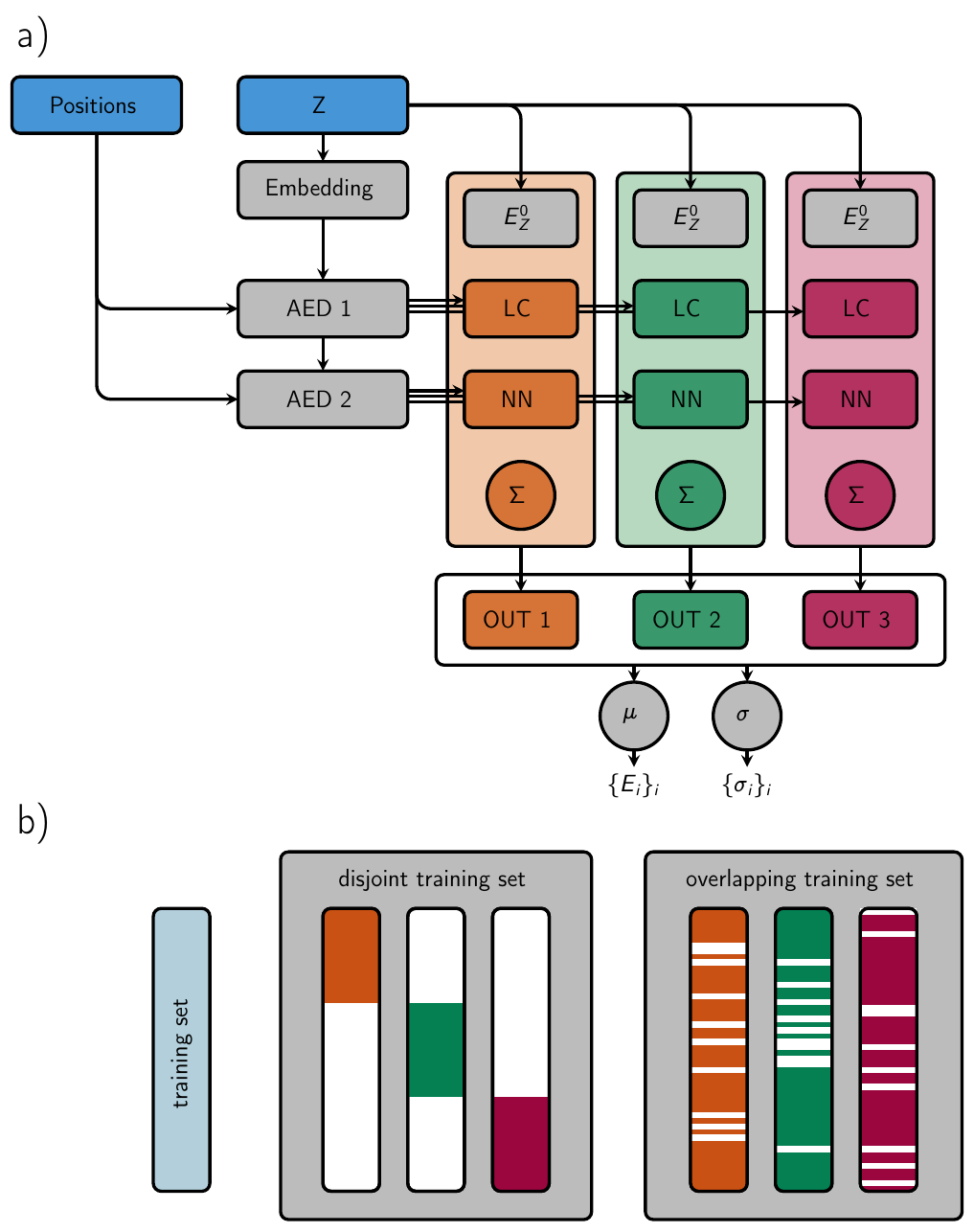}
\caption{\label{fig:architecture}
Panel a) shows a schematic description of MHC architecture, for an example model with two MACE layers and three committee members.
Panel b) illustrates two different options of splitting the training data across the heads: ``disjoint'' and ``overlapping''.
}
\end{figure}

As shown in Figure~\ref{fig:architecture}, we construct neural network committees for uncertainty quantification by attaching multiple readout heads to message passing node features and use their disagreement as an uncertainty metric.
Multiple-readout heads have been used to train on multiple datasets and simplify fine-tuning~\cite{Batatia2023/10.48550/arXiv.2401.00096}.

The geometric message passing layers of MACE construct atomic environment descriptors $\vect{h}_i^{(l)}$ for each atom $i$, and layer $l$.
We attach separate layer dependent readout heads $\mathcal{R}$,
\begin{equation}
    E_{i,n}^{(l)} = \mathcal{R}_{l,n} \left( \vect{h}_i^{(l)} \right),
\end{equation}
where n indexes the different committees.
As in the MACE architecture, the first layers have linear readouts, while the readout of the last layer is a multilayer perceptron.

The total energy for committee member $n$ is obtained by summing over all atoms, incorporating both the contributions from the readouts and the isolated atom energies,
\begin{align}
    E_n = \sum_i E_{i,n} = \sum_i \left( E_{Z(i)}^{(0)} + \sum_l E_{i,n}^{(l)} \right),
\end{align}
where $E_{Z(i)}^{(0)}$ denotes the isolated atom energy of species $Z(i)$.
The forces $\vect{F}_{i,n}$ are calculated as the negative gradient of this total potential energy with respect to the Cartesian positions.

The committee prediction is then obtained by taking the average over all committee members,
\begin{align}
    E &= \frac{1}{A} \sum_a E_a \quad
    \vect{F}_i = \frac{1}{N} \sum_n^N \vect{F}_{i,n},
\end{align}
and the uncertainty estimation is the standard deviation of the energies or forces.
To obtain an uncertainty estimate for each atom in the system, we take the average standard deviation over the three force components $\alpha$, such that
\begin{align}
    \sigma_{F,i} = \frac{1}{3} \sum_{\alpha} \left( \frac{1}{N} \sum_n^N (F_{i,n}^{(\alpha)} - F_i^{(\alpha)}) \right)^{\frac{1}{2}},
\end{align}
which can be used as the error estimate.

\subsection{Distribution of training data}
During training, each structure gets assigned to a specific head and is subsequently used to optimize the AED and the assigned head.
If we want to use a structure for multiple heads, it will appear multiple times in the training set with different assigned heads.
To increase the heterogeneity between the heads, we train each of them on a different subset of the complete training set.
We consider two possible strategies: either randomly sampling a fraction of the total set of structures for every head, or evenly distributing the whole dataset across the heads without any overlap between the subsets.
Both strategies are illustrated in panel b) of Figure~\ref{fig:architecture} and bring different benefits.
The first strategy, in which a fraction of the total set of structures is randomly sampled for each head, is called ``overlapping'' due to the overlaps between resulting training sets.
Its potential downside is that in the final concatenated training set, on which the common parts of the model are trained, structures appear multiple times without an even distribution among them.
Therefore, the AEDs will be trained more often on some structures than others.
The second strategy, in which the whole dataset is evenly distributed among the heads, is called ``disjoint'' as the training subsets do not overlap.
As will be shown in the Results section, there can be situations where the datasets used to train the output heads of the model are too sparse, and the prediction accuracy consequently decreases significantly if the original dataset is already extremely small.
The main benefit of the disjoint strategy is that it increases the diversity between the output heads more strongly, mitigating known problems of a common bias between the committee members~\cite{Kahle2022/10.1103/PhysRevE.105.015311}.

\subsection{Uncertainty scaling}
Multiplying the committee disagreement with a uniform scaling factor calculated from an independent validation set can correct for an underestimation of the true uncertainty due to biases inherent to the models~\cite{Imbalzano2021/10.1063/5.0036522}.
\begin{align}
    \alpha^2 = \frac{1}{N_{\mathrm{val}}} \sum_{i \in \mathrm{val}} \frac{(\Delta y_i)^2}{\sigma_i^2},
\end{align}
where $\Delta y_i$ and $\sigma_i$ are the error and committee disagreement of the i-th element of the validation set.
We note the scaling factor wherever it was applied.

\subsection{Pearson correlation coefficient}
The Pearson correlation coefficient~\cite{Pearson1895/10.1098/rspl.1895.0041}
\begin{align}
    r(\epsilon, \sigma) = \frac{
        \sum_{A} \left( \epsilon(A) - \overline{\epsilon} \right) \left( \sigma(A) - \overline{\sigma} \right)
    }
    {
        \sqrt{\sum_A \left( \epsilon(A) - \overline{\epsilon} \right)^2}
        \sqrt{\sum_A \left( \sigma(A) - \overline{\sigma} \right)^2}
    }
\end{align}
is a measure of correlation between two datasets, in our instance, the error $\epsilon$ and the uncertainty $\sigma$.
In the case of energy predictions, this uses the energy error and uncertainty per structure, and in the case of force predictions, it uses the root mean square error and mean uncertainty of all three force components of one atom.
$\overline{x}$ denotes the mean over the whole dataset.
The value of the correlation coefficient can range from -1 to 1, corresponding to perfect anti-correlation and perfect correlation, respectively, and 0 meaning no correlation at all.

\subsection{Implementation}
We implemented MHCs in MACE 0.3.7 and used this version of the code for the whole project.
The output heads are configured to calculate the potential energy predictions of all heads simultaneously.
Thus, the mean and standard deviation of the committee energy can be obtained with essentially no additional computational costs.
In contrast, due to the intrinsic limitations of automatic differentiation, a single backward pass can not yield the forces for all the heads, and thus their standard deviation.
One can still calculate the average forces across the committee heads at the same costs as a normal model, but the standard deviation requires multiple reverse passes through the computational graph, adding some computational overhead.

\subsection{Training Settings}
The custom-built models consisted of two message-passing layers, with 32 channels and a maximum tensor order of $l=1$.
The multi-layer perceptron in the output blocks of the final layer had 16 nodes in the hidden layer of each output block and used the default SiLU gate.
The radial cutoff was set to \SI{6.0}{\AA}, resulting in an effective field of view of \SI{12.0}{\AA}.
The isolated atom energies were always set to the ones specified in the respective datasets.
The compositional differences between the committee types lead to different sizes of the full training dataset used for each model.
To keep the total number of optimization steps consistent between the different types of models, we adapted the training parameters in our setups.
This also means that each member of the naive committee received as many optimization steps as the MHC.
When investigating different training set sizes, we again ensured that the number of optimization steps remains constant for all training sets.
A weighted mean squared error with a force-to-energy weight ratio of 100 to 1 was used for model optimization.
For the final 25\% of the training, the Stochastic Weight Averaging approach with default settings was used to further optimize the energy predictions.
The 3BPA models were trained for 5000 optimization steps, the water models for 2200 steps, and the rMD17 models for 16,000 steps.
The output blocks of the adapted foundation model were trained with the same basic parameters for a total of 800,000 steps.
To condense the MPtrj dataset, we used an iterative QbC workflow~\cite{Schran2020/10.1063/5.0016004}.
In each iteration, we sorted the structures based on the maximum disagreement of their force components and added the 100 highest ranking structures to the training set.
Afterward, the output heads were retrained for a reduced number of optimization steps.

\section{Results}
In this section, we demonstrate the power of the MHC methodology by applying it to a series of systems and models of increasing complexity.
We start with a gas-phase molecule, move to the condensed phase with liquid water, explore chemical space with a model for multiple organic molecules, and finally enhance a foundation model with uncertainty prediction.
For the upcoming results, it should be noted that in every calculation, the model's hyperparameters can influence the results in many unintended, subtle ways, for example the shape of the correlation distribution.
We have kept the hyperparameters as consistent as possible between models to minimize these factors.

\subsection{3BPA}
\begin{figure*}
\centering
\includegraphics[width=\linewidth]{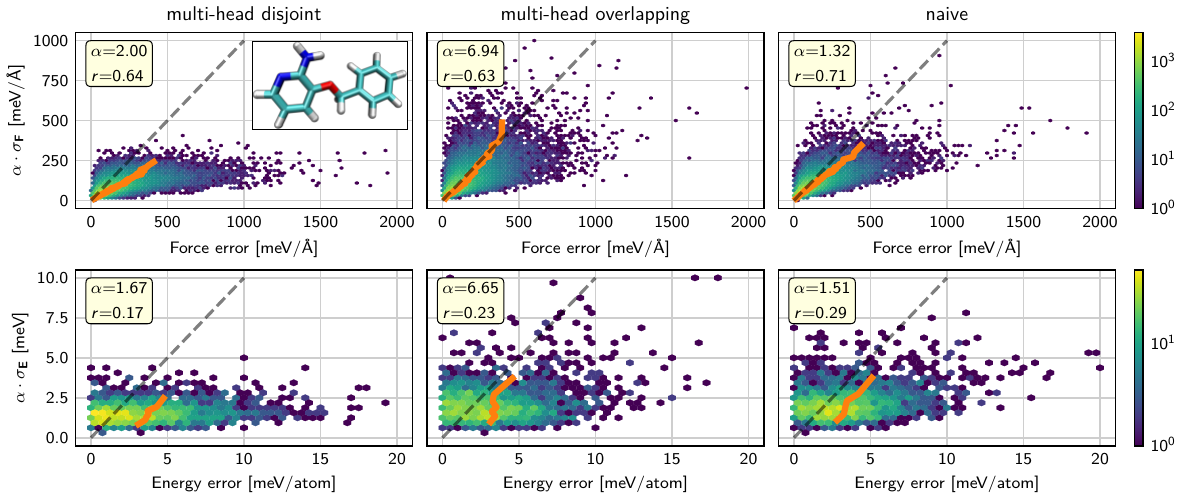}
\caption{\label{fig:3BPA}
The correlation between the RMSEs of the models and the scaled committee disagreements for forces per atom and energy of the whole molecule.
The scaling parameter $\alpha$ and the Pearson correlation coefficient is given in the top left corner of each subplot.
The orange line indicates the binned average RMSE for all atoms or structures for forces and energies, respectively, binned along the committee disagreement.
All bins contain the same number of data points and are, therefore, not of equal size.
The gray line shows perfect correlation between uncertainty measure and actual error.
The 3-(benzyloxy)pyridin-2-amine (3BPA) molecule used in this analysis is shown in the inset of the first plot.
}
\end{figure*}

As a first step, we show that the MHC can be used to estimate the uncertainty of MLP predictions.
To this end, we test and compare the two different options for subsampling the MHC training set.
We then compare these against the baseline of a naive committee, comprising individual full MACE models each trained on randomly chosen subsets of the training data.
The aim for the MHC disagreement is to be as close to the naive committee disagreement as possible, despite sharing most of the network's trainable weights and therefore having less room for divergence between the committee members.
For these first tests, we use the 3BPA (3-(benzyloxy)pyridin-2-amine) dataset~\cite{Kovacs2021/10.1021/acs.jctc.1c00647}, which contains structures of the drug-like 3BPA molecule from MD trajectories at different temperatures.
In the 3BPA molecule, shown in the top left panel of Figure \ref{fig:3BPA}, the most important degrees of freedom are the torsions along the bonds connecting the two six-membered rings.
The training data is based exclusively on \SI{300}{K} structures, whereas the test data originates from sampling at \SI{1200}{K} and therefore reaches a broader region of configuration space than the training set.
Figure~\ref{fig:3BPA} shows the correlation between the actual error in force or energy per atom and the corresponding standard deviation of the committee predictions.
The distribution of the values is shown as a two-dimensional histogram using hexagonal bins on a logarithmic scale.
In order to map to the true generalization error, these standard deviations were scaled in postprocessing using a scaling factor calculated from the validation set~\cite{Imbalzano2021/10.1063/5.0036522}, as detailed in the Methods section.
The individual scaling factors $\alpha$ are given in the upper left corner of every panel of Figure~\ref{fig:3BPA}.
To better illustrate the overall trend, we also bin the distribution along the uncertainty axis to contain an equal number of data points in each bin and calculate the mean error for each bin, shown as orange lines.
A reference gray line indicates perfect correlation between errors and uncertainty.
In an attempt to quantify the correlation, we calculated the Pearson correlation coefficient~\cite{Pearson1895/10.1098/rspl.1895.0041} $r$, which can be found in each panel below the scaling factor.
We see a significant spread of the distribution surrounding the ideal line, while the binned averages follow the optimal correlation closely.
For all models investigated, there are data points with high uncertainty but small error, which is not an issue, as a model can produce an accurate prediction ``by accident'', despite high uncertainty.
Importantly, there are no instances where the model's error is high while the uncertainty is low, which would be a clear indication of an unreliable uncertainty estimator.
For all of our committee models, there is a clearly empty zone in the bottom right corner, especially for forces, which means that an increased error will always be indicated by an increased uncertainty.

Furthermore, the MHCs perform almost equally well as the naive committee, indicating that even for a shared description of the atomic environment, the flexibility of the output heads is sufficient to obtain a meaningful committee disagreement.
Unfortunately, for all three committees, the uncertainty correlation is worse for energy than for forces.
A correlation is not visually apparent from the distribution of values, and only the binned mean uncertainty reveals a general trend of increasing uncertainty with increasing error.
The clear zone of high error at low disagreement is less pronounced than for the forces, as the committee is over-confident for structures with a high error.
However, this problem exists for all three committee types, and the MHC does not perform worse than the naive committee.
The scaling factor for the naive committee is lowest for both properties, followed by the MHC with a disjoint training set.
The scaling parameter for the model, where the subsets of the heads overlap, is significantly higher.
This result is repeatedly observed across every system we investigated.
However, after scaling, all three models exhibit similar behaviour, rendering this observation less consequential.

\begin{figure}
\centering
\includegraphics[width=\linewidth]{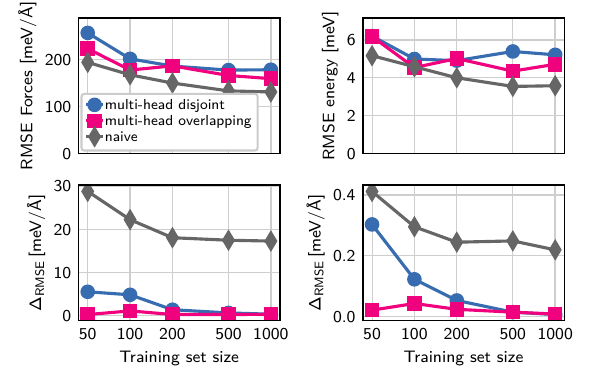}
\caption{\label{fig:error_evolution}
Errors of different committee architectures on the 3BPA@1200K test set as a function of the size of the training set.
In the left column the force RMSE and in the right column the energy RMSE is shown.
The top row displays the error of the whole committee, whereas the bottom row displays the difference between the committee error and the average error of the individual committee members.
Please note, that the scaling of the x-axis is logarithmic.
}
\end{figure}

Finally, we use the 3BPA system to analyze the prediction accuracy of the different model types.
The evolution of the force and energy RMSEs shown in Figure~\ref{fig:error_evolution} illustrates that the naive committee consistently outperforms the MHCs in prediction accuracy.
However, this is expected, as the naive committee has almost 8 times as many trainable parameters as the MHCs due to the full independence of every committee member.
Comparing the two different versions of MHCs shows a slight advantage in terms of accuracy for the one with overlapping training sets.
In this version, each multi-head training set consists of 80\% of the training data for each head for the overlapping committee and just 1/8 of the set for the disjoint committee.
The training set for each head comprises 80\% of the training data for the overlapping committee and just 12.5\% of the training data for the disjoint committee.
We conclude that the smaller size of the dataset available to each head in the disjoint committee leads to this small discrepancy.
In the bottom row of Figure~\ref{fig:error_evolution}, the difference between the error of the whole committee and the average error of the committee members is shown.
For the naive committee, the committee consistently outperforms the individual members due to the increased number of trainable parameters.
For the overlapping committee, the differences are small and independent of the number of training structures.
The improvement in performance typically associated with committee MLPs remains absent, as the MHC is conceptually similar to adding a dropout layer to the output blocks.
While classic dropout layers set the output of random nodes to zero during training, the MHC does so for all nodes connected to certain output heads.
For every structure in the training set, the same nodes are consistently removed in every epoch of the training process.
It would be unreasonable to expect a significant performance boost from combining multiple of such predictions.
Therefore, no significant improvement in predictive power should be expected.
Contrary to this expectation, the disjoint model shows a noticeable improvement when introducing the committee.
This improvement stems from the number of training examples each output head sees during training.
Especially for small training datasets, the subsets for each head are extremely sparse, negatively affecting the prediction accuracy.
With a large overall training set size, this sparsity decreases, leading also to more similar single and committee predictions.

\subsection{Liquid water}
\begin{figure}
\centering
\includegraphics[width=\linewidth]{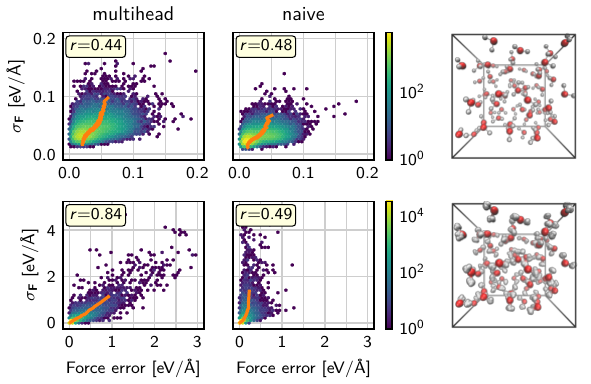}
\caption{\label{fig:water}
Correlation of committee force disagreement and RMSE for each atom in a 64-molecule box of liquid water.
The top row shows results for structures with classical nuclei --- the same distribution as the training data.
The bottom row shows results for structures with quantum nuclei --- a distribution different from the training data.
In each panel, the Pearson correlation coefficient $r$ is given in the top right corner.
}
\end{figure}

Moving from the gas phase to the condensed phase, we next test MHCs on bulk liquid water.
Here, we choose to focus on the disjoint sub-sampling of the training set, because the uncertainty measure resembles that of the naive committee more closely --- especially with respect to the calibration factor.
We fit both naive and disjoint committees to a previously published training set of 111 structures of liquid water, each containing 64 molecules~\cite{Schran2020/10.1063/5.0016004}.
We tested the force uncertainty prediction of the models using 500 structures with classical nuclei at \SI{300}{K} (in-distribution performance), and using 500 structures with quantum nuclei at \SI{300}{K} (out-of-distribution performance), as shown in the top and bottom row of Figure \ref{fig:water}, respectively.
The forces for the classical MD data are predicted very well, and, therefore, the uncertainties are low.
Both committees feature much larger force errors for the path-integral structures.
In this case, the MHC results in a stronger correlation between committee disagreement and forces but also more instances of a high force error for certain atoms than the naive committee.
It is important to note that the plot style in Figure~\ref{fig:water} places a strong emphasis on the outliers of the distribution, while the bulk of the distributions are in the area of low errors and low uncertainty.
Nevertheless, it is evident that in the prediction of uncertainties, the MHC is on par with the naive committee.

\subsection{rMD17}

\begin{figure*}
\centering
\includegraphics[width=\linewidth]{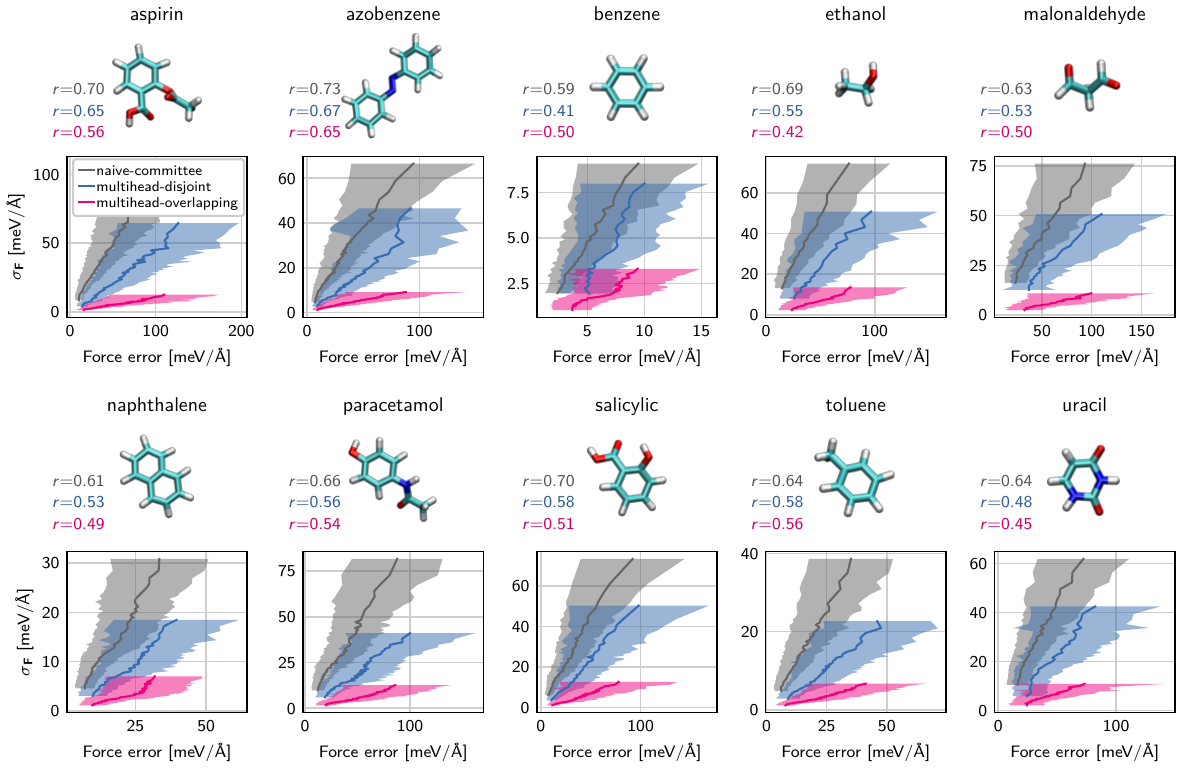}
\caption{\label{fig:rMD17}
Correlation between the unscaled committee disagreement and the force RMSE for individual atoms in the systems.
The curves are plotted individually for all 10 different molecules in the rMD17 dataset and the structure of the respective molecule is shown above the plots.
The curves show the average force RMSE binned along the committee disagreement, as the orange curves in Figure \ref{fig:3BPA} and \ref{fig:water}.
The shaded areas around the lines are the standard deviations within each bin.
The Pearson correlation coefficient $r$ is given above the correlation plots in the color of the respective model.
}
\end{figure*}

While each of the previous test cases focused on a single molecular system, in this section, we expand our investigation to cover a more diverse set of molecules within one model to probe uncertainty predictions across chemical compound space.
To test this, we employ the rMD17~\cite{Christensen2020/10.1088/2632-2153/abba6f} dataset comprising MD structures sampled at \SI{500}{K} of 10 different organic molecules (consisting of H, C, N, and O atoms), which are shown in Figure~\ref{fig:rMD17}.
We randomly selected 50 structures from each of the 10 molecules to form our training set and 1000 structures of each molecule for separate per-species test sets.
The unscaled mean and standard deviation of the uncertainties are shown in Figure~\ref{fig:rMD17} against the force errors for the multi-head and naive committees for all 10 molecules individually, to show how the model can handle predictions of different complexities.
For an easier comparison, we plot the binned averages of all three approaches in one subplot.
The shaded area indicates the standard deviation of the data in each bin to illustrate the spread of the data.
The magnitudes of the uncertainty are dependent on the molecule under investigation, but the general trends are very consistent, as are the comparisons between the committee types.
Analogously to our previous findings, the scales of the three curves differ substantially, while the correlations remain fairly similar for all three committee types.
As expected, the naive committee shows both the highest level of uncertainty and the highest Pearson correlation coefficient due to the large differences between its members, and the overlapping MHC shows the lowest, as its members are the least diverse.
We also conducted a correlation analysis using the relative log-likelihood method~\cite{Kellner2024/10.1088/2632-2153/ad594a}, coming to the same conclusions (details in the SI, Section~\ref{si-sec:corr-coef}.
Unfortunately, the correlation between energy uncertainty and energy errors is substantially weaker.
The Pearson correlation coefficient calculated over the full test set is only 0.37 for both the naive committee and the disjoint MHC and 0.29 for the overlapping MHC.
We discuss the energy uncertainty in further detail in the SI, Section~\ref{si-sec:energy}.
We also use the rMD17 dataset to examine committee disagreement for unknown atomic systems, as this will be relevant in the context of foundation models.
Therefore, we remove one molecule from the training data and train the committees on the remaining 9 molecules.
When we examine the committee disagreement on a test set of the molecule that was taken out of the training data, we find that the uncertainty still correlates well with the errors but can be distinctively low for certain atoms.
We attribute this to the local environment of these atoms, which is similar to the local environment of atoms present in the training data.
We discuss these results in more detail in the SI~\ref{si-sec:missing_molecules}.
Importantly, we find that our previous conclusions are also valid for heterogeneous datasets and that committee disagreement works well with large, diverse training sets and out-of-domain test structures.

\subsection{MACE MP-0 foundation model}

After showing that the MHC provides a reliable measure of prediction uncertainty, we adapt it to a foundation model.
We start from the MACE-MP-0b foundation model~\cite{Batatia2023/10.48550/arXiv.2401.00096} trained on the MPtrj dataset~\cite{Deng2023/10.1038/s42256-023-00716-3}, which contains 146k crystalline structures calculated using density functional theory at the PBE+U~\cite{Perdew1996/10.1103/PhysRevLett.77.3865} level.
We then equip this foundation model with an MHC by adding eight new output heads with a random initialization of weights to the existing pre-trained head.
As with custom-build models, the committee prediction is the mean of all new output heads, excluding the original head.
When training the MHC output heads, we leave the weights of the AED block and the original head fixed.
To obtain a more compact training set for the MHC, we use an iterative QbC workflow based on its committee disagreement~\cite{Schran2020/10.1063/5.0016004}.
This reduces the original dataset to 8,000 structures --- just 5\% of its original size.
This condensed training dataset is divided across the eight committee heads using the ``disjoint'' distribution, resulting in 1,000 training structures for each head.
By using a much smaller training set and taking advantage of the pre-trained model, the computational cost of training is greatly reduced.
Overall, we are adding roughly 15,000 parameters to the model, accounting for less than 0.2\% of the total model size.
It took 37 hours on an NVIDIA Hopper H100 GPU to execute the 800,000 optimization steps of the whole MHC, a small fraction of the 2,600 GPU hours of the original model (trained on NVIDIA A100 GPUs across multiple nodes)~\cite{Batatia2023/10.48550/arXiv.2401.00096}.
Especially for the smaller foundation models, this training effort can also be performed on consumer-grade GPUs.

\begin{figure}
\centering
\includegraphics[width=\linewidth]{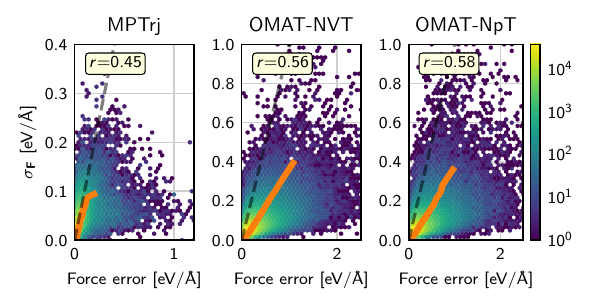}
\caption{\label{fig:mp0_correlation}
Correlation between the committee disagreement and the force error for the modified MACE MP-0b model, where the output heads were trained on a reduced sample of the MPtrj dataset containing 8000 structures.
The first subplot is on test data from the MPtrj dataset, which is not included in the reduced dataset.
For the two subsequent plots, 10,000 structures from the NVT and NpT dataset of the OMAT database were used.
As in the previous figures, the orange lines indicate the average force RMSE, binned along the direction of the uncertainty measure.
The Pearson correlation coefficient $r$ is given in the top left corner of each panel.
}
\end{figure}

For evaluation, we used 10,000 out-of-distribution structures taken from the NVT and NpT OMAT datasets~\cite{Barroso-Luque/10.48550/arXiv.2410.12771}, which are calculated with the same reference method as the MPtrj dataset, and 10,000 structures from the MPtrj dataset not selected during QbC.
For the OMAT datasets, a very small number of predictions had an extremely high error for all models, including the original MACE-MP-0 model.
Due to the nature of the Pearson correlation coefficient and the root mean square errors, these few predictions dominate the final results.
Therefore, we decided to exclude all force components for which the error of the original MACE-MP-0 model is higher than \SI{5}{eV/\AA} and all energy predictions with an energy error higher than \SI{1}{eV}.
This amounts to a total of 18 energy values and 764 force components, or 0.019\% of the combined OMAT test sets.
Figure~\ref{fig:mp0_correlation} shows the correlation between the actual force errors per atom and the uncertainty prediction for the three datasets in question.
Compared to previous cases, the results are more spread out, as both the training and test data are much more diverse.
In the case of OMAT, this means that many of the systems in the test set are not present in the training data at all.
The MPtrj dataset, on the other hand, functions as an indicator for the in-data regime, as it is not independent of the training data of the model.
The AEDs of the model were trained on the full MPtrj dataset, including the structures of this test set.
Overall, the committee disagreement correlates well with the error of the committee prediction and therefore functions as a reliable uncertainty measure even for foundation models.
However, unlike with custom-made models, we observe occasional instances where a low uncertainty coincides with a high error.
This is most likely due to structures in the test set, which have similar characteristics to some structures in the training set, resulting in uniform predictions by the committee.
This also explains why the Pearson correlation coefficient is moderately lower than for custom-made models.
In the MPtrj dataset, where the structures of the test set were used to train the original foundation model, but not the committee output heads, this effect is particularly pronounced.
However, overall, these instances are isolated enough not to deteriorate the reliability of the MHC uncertainty prediction for foundation models.

\begin{figure}
\centering
\includegraphics[width=\linewidth]{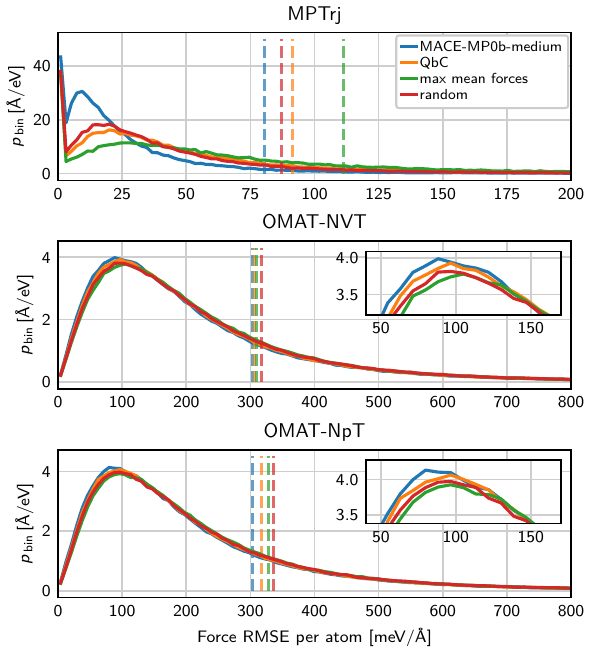}
\caption{\label{fig:mp0_distribution}
The distribution of force errors for different versions of the MACE MP0b foundation model.
The datasets are the same as in Figure \ref{fig:mp0_correlation}.
The blue curves correspond to the original model.
For the remaining models, the original output head was replaced by a multi-head output module and retrained on 8,000 structures sampled from the MPtrj database.
For the orange line, the reduced training set was sampled using QbC, for the green line the structures with the highest mean force components were selected, and the red line corresponds to a random sampling.
The vertical dashed lines indicate the RMSE of the complete dataset for each model.
For increased clarity, a inset showing the peaks of the error distributions at a higher resolution were added for the two OMAT datasets.
}
\end{figure}

An important question that remains is whether the new committee model predictions are still accurate compared to the original, despite being trained on a small fraction of the original dataset.
To test whether QbC is advantageous compared to other, computationally cheaper options, we created two alternative training sets.
During the QbC run to select the reduced training dataset, it is noticeable that structures with high force components are selected at a much higher rate than structures close to the equilibrium.
Therefore, we created a training set of the 8,000 structures with the highest force per atom.
Additionally, we also randomly selected 8,000 structures from the initial dataset to form a third training set.
Both datasets were used to train MHC models in the same way as the QbC dataset.
Figure~\ref{fig:mp0_distribution} shows the distribution of the force errors for the original MACE-MP-0 model and the new MHC models.
As expected, the original MACE-MP-0 model has the highest prediction accuracy of all models, but the advantage over the re-trained models is small.
Among the models trained on less data, the model with the QbC selected data performs the best, but the differences compared to other models are modest.
For the MPtrj dataset, the original MACE-MP-0 model has a much stronger advantage over the retrained models than for the OMAT data.
This is because the OMAT data is entirely unseen by all models, whereas the MPtrj test set was part of the MACE-MP-0 training dataset, and therefore, this strong performance should be expected.
This also explains why the randomly selected subset outperforms the QbC selected set, as we sampled it from the structures not picked during QbC.
Therefore, some structures of the randomly selected training set are included in this test set.
As we show in the SI~\ref{si-sec:1000_structures}, condensing the training set even further will eventually lead to a measurable decrease in prediction accuracy.

As an alternative to the MHC, one could make use of the existing fine-tuning mechanism and form a naive committee of independently fine-tuned foundation models.
We test this approach in a set-up that is consistent  with the MHC in the number of available training structures, as well as the required computational resources.
As detailed in the supporting information (Section ~\ref{si-sec:finetuned_committee}), we find that the correlation between uncertainty and error remains adequate, but the prediction error of the committee of fine-tuned models roughly quadruples compared to the original foundation model.
This further underscores both the prowess and the resource efficiency of the MHC approach for foundation models.

Overall, we conclude that our approach of adding an MHC to a foundation model preserves its predictive power.
This is, on one hand, achieved by leaving the original foundation model --- including its output head --- intact.
On the other hand, we have shown that the added output heads, trained on a condensed training dataset, also largely retain the foundation model's performance, as most of the predictive capacity lies within its AEDs, which were extensively trained on the full MPtrj dataset.
Optimizing the weights of the AEDs further on a condensed training set would degrade the model's performance.

\section{Conclusions}

In this work, we have developed a method to build committee MLPs using the MACE graph neural network potential.
The AEDs are shared between committee members, and the committee is formed by attaching multiple output blocks to the descriptor layers.
This allows us to build a committee that is more efficient during training and prediction. %
The main benefit of this committee over regular MACE models is that the standard deviation of the committee's predictions can function as an estimation for the uncertainty of the model.
When testing this uncertainty estimation with committee MLPs trained on established datasets such as 3BPA, water, and rMD17, we found that for forces, there is a strong correlation between the committee disagreement and the actual prediction error.
In particular, the MHC rarely featured instances where the error was high despite low uncertainty.
Unfortunately, the correlation is considerably weaker for energy predictions.
Crucially, we saw no drop in performance when comparing the uncertainty estimation of the MHC with that of naive committees with completely independent committee members.
The output modules on their own provided enough flexibility and diversity for a reliable uncertainty estimate.
Furthermore, we compared two different strategies to distribute the full training set across the output heads.
One, where the training set was split evenly between the heads, and one, which allowed for overlap between the subsets and used independent randomly sampled training sets for each head.
Both methods displayed a similar level of correlation, but the disjoint model showed preferable scaling of the uncertainty.

The second part of this work used the MHC in active learning to condense a reduced training set out of the original MPtrj dataset, a widely used training set for foundation models.
We selected 8000 out of 146k structures from the MPtrj dataset and used them to train the output heads of an adapted medium-sized MACE-MP-Ob foundation model.
The AEDs of the model were left untouched to preserve its expressive capabilities.
We showed that the MHC based on foundation models displays a good correlation between committee disagreement and actual force error, even though instances where the uncertainty is underestimated are slightly more common.
When comparing the predictions of the original MACE-MP-0b model with the new model, we saw only a small drop in performance, although the output heads were trained on only 5.5\% of the full training dataset.
We also examined the criteria for selecting the reduced training sets and found that the QbC-selected model performs the best, even though the advantage is moderate.
Given the compact nature of the condensed data sets, this opens up the possibility to obtain foundation models trained on expensive high-level electronic structure methods by recalculating the previously condensed training set and optimizing only the output blocks.
QbC can help ensure that the performance stays as close as possible to the original foundation model.
Overall, the uncertainty estimation of the MHC architecture introduced here further increases the robustness of foundation models.

\begin{acknowledgments}
The authors thank Ilyes Batatia and Gábor Csányi for valuable discussions.
H.B. and P.S. acknowledge support from the Charles University Grant Agency, project number 248923, and the International Max Planck Research School for Quantum Dynamics and Control.
L.L.S. would like to acknowledge support from the UKRI Critical Mass grant, project reference EP/V062654/1, the Isaac Newton Trust, award number G122390 and Wolfson College, Cambridge.
O.M. acknowledges support from the Czech Science Foundation, project No. 21-27987S.
C.S. acknowledges financial support from the Royal Society, grant number RGS/R2/242614.
\end{acknowledgments}

\section*{Conflict of Interest Statment}
The authors have no conflicts to disclose.

\section*{Author contributions}
H.B., O.M., and C.S. conceived of the idea.
H.B. developed and implemented the method, carried out most of the calculations, and wrote the initial draft of the manuscript.
P.S. supported the training of the foundation models.
All authors contributed to the interpretation of the results and writing of the manuscript.

\section*{Data availability}
The data that support the findings of this study are openly available in Zenodo at http://doi.org/[doi] (doi will be added when published)

\section*{Code availability}
The MACE implementation used for all simulations can be found in this pull request: \url{https://github.com/ACEsuit/mace/pull/800}.

\section*{References}

\putbib

\end{bibunit}

\clearpage

\setcounter{section}{0}
\setcounter{equation}{0}
\setcounter{figure}{0}
\setcounter{table}{0}
\setcounter{page}{1}

\renewcommand{\thesection}{S\arabic{section}}
\renewcommand{\theequation}{S\arabic{equation}}
\renewcommand{\thefigure}{S\arabic{figure}}
\renewcommand{\thepage}{S\arabic{page}}
\renewcommand{\citenumfont}[1]{S#1}
\renewcommand{\bibnumfmt}[1]{$^{\rm{S#1}}$}
\newcommand{\NLL}{\mathrm{NLL}}
\title{Supporting information for: \mytitle}
{\maketitle}

\onecolumngrid
\fontsize{12}{14}\selectfont

\begin{bibunit}

\linenumbers\relax

\section{Energy disagreement in rMD17}
\label{si-sec:energy}
\begin{figure*}
\centering
\includegraphics[width=\linewidth]{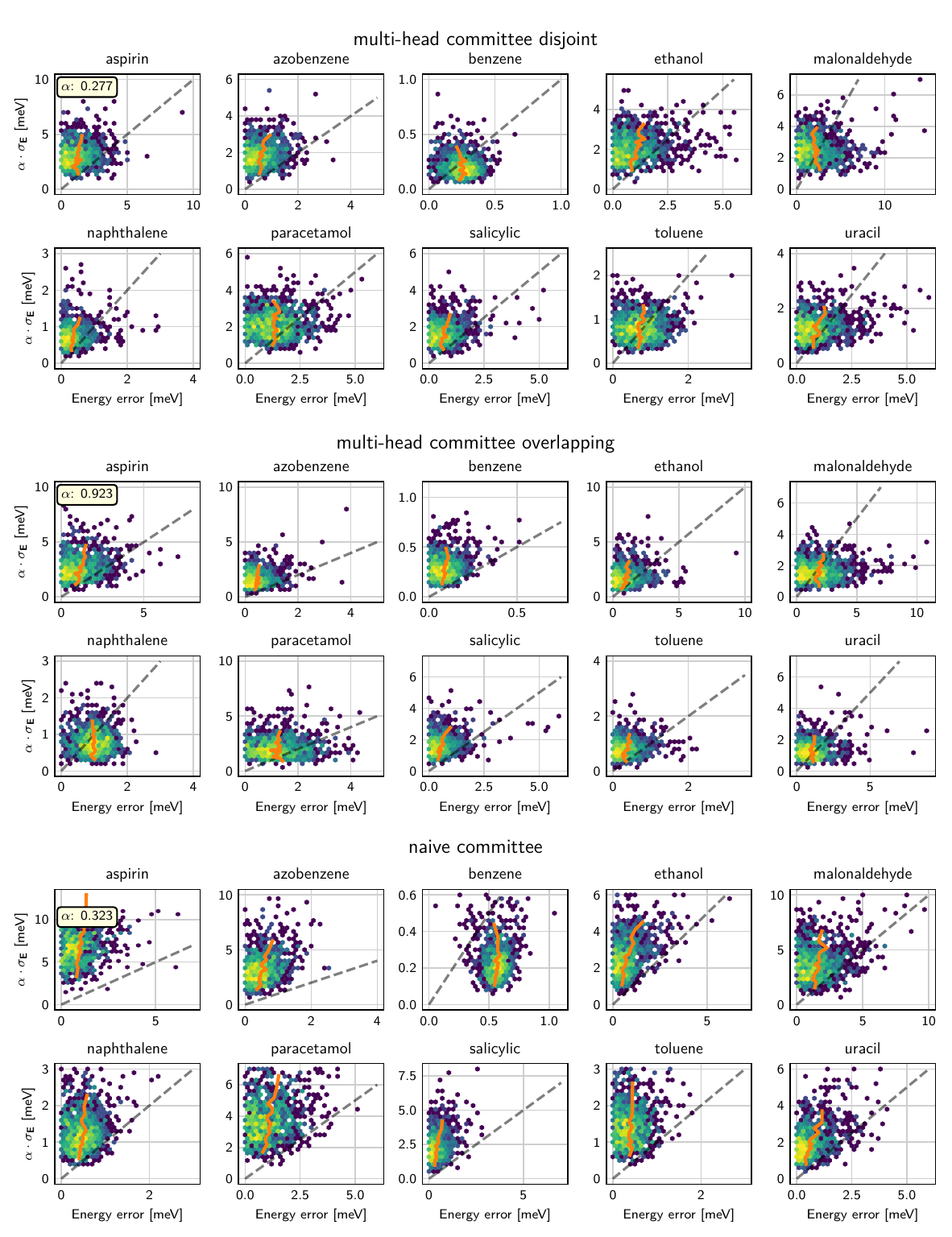}
\caption{\label{si-fig:energy-rmd17}
This figure shows the correlation between the energy committee disagreement and the energy error.
The figure consists of three blocks, the top and middle blocks for the multi-head committee trained on a disjoint and overlapping datasets respectively, and the bottom block for the naive committee.
Each block consists of 10 plots for the 10 different molecules in the rMD17 dataset.
The orange line shows the binned averages of the energy error and the grey line indicates perfect correlation.
}
\end{figure*}

Figure \ref{si-fig:energy-rmd17} shows that for all types of committees we tested, the correlation between the energy committee disagreement and the true errors is low for most systems in the rMD17 dataset~\cite{Christensen2020/10.1088/2632-2153/abba6f}.
Plotting the mean energy errors binned along the uncertainty in an orange line shows a trend of increasing uncertainty at high errors in some cases, but the trend is often weak and usually far from the optimal correlation indicated by the grey dashed line.
However, the important criterion of having few instances of high error despite low uncertainty is still met, especially for the multi-head committee with a disjoint training set and the naive committee.
Therefore, the energy uncertainty can still be valuable for on-the-fly tracking of uncertainty.
Unfortunately, for the overlapping variant of the multi-head committee, the number of problematic underestimations of the uncertainty increases.
Overall, the uncertainty estimate is substantially worse than the force disagreement on the same dataset or the energy disagreement on a smaller dataset such as 3BPA (see Figures \ref{fig:rMD17} and \ref{fig:3BPA} of the main article respectively).

\section{Correlation coefficients and relative log likelihood}
\label{si-sec:corr-coef}
\begin{figure*}
\centering
\includegraphics[width=.5\linewidth]{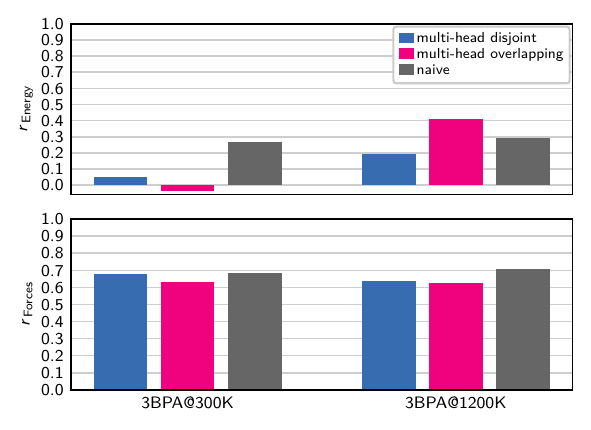}
\caption{\label{si-fig:cc-3bpa}
The Pearson correlation coefficients between errors and committee disagreement for the 3BPA datasets.
The top panel shows the correlation coefficient for energy predictions and the bottom panel for force predictions.
The left side of each panel depicts the coefficient for a test set taken from an MD trajectory at 300~K, which is the same as the training set.
The right side shows results from the same test set as was used in the main article, which comprises structures taken from a trajectory at 1200~K.
}
\end{figure*}

\begin{figure*}
\centering
\includegraphics[width=\linewidth]{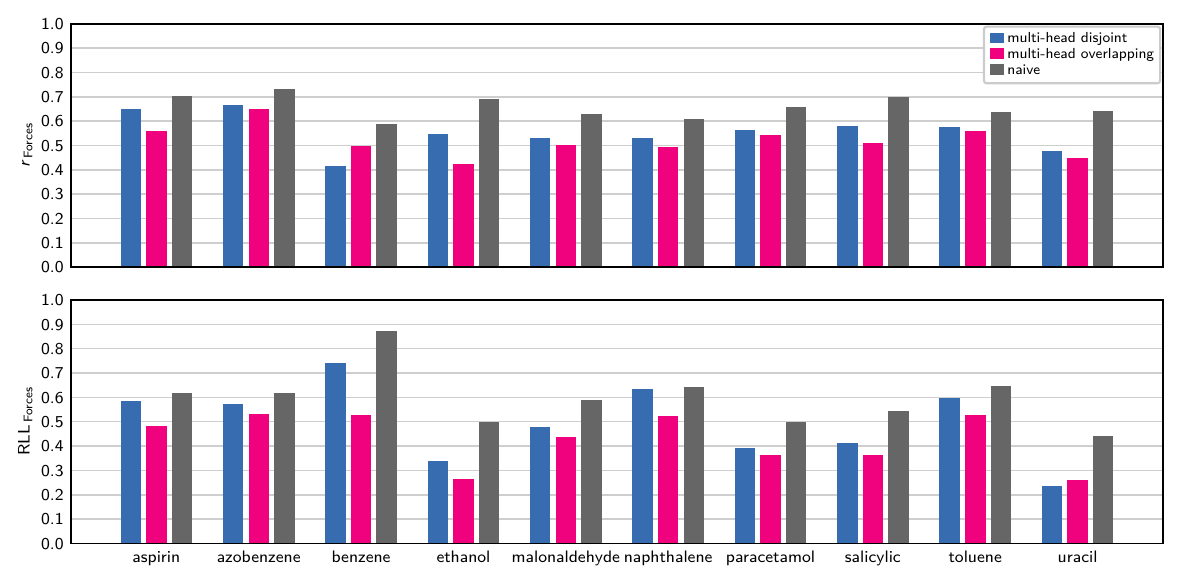}
\caption{\label{si-fig:rll-cc-rmd17}
The Pearson correlation coefficient (top) and relative log-likelihood (bottom) of the force committee disagreement compared to force errors on the rMD17 dataset.
The results are split into ten blocks for the ten molecules in the dataset.
}
\end{figure*}

When investigating how well committee disagreement functions as an uncertainty estimation for the true error in the main article, we focused on a graphical evidence and plotted the uncertainty estimate against the actual error.
However, there are alternatives to assess uncertainty in a more quantitative way, which we show here.
The first is the Pearson correlation coefficient~\cite{Pearson1895/10.1098/rspl.1895.0041}, which we explained in the Section \ref{sec:methods} of the main paper.
For many results in the main paper, the Pearson correlation coefficient was given as an annotation in the figures.
The second option, which we are exploring, is the relative log-likelihood (RLL), proposed by Kellner et al.~\cite{Kellner2024/10.1088/2632-2153/ad594a}:
\begin{align}
    \mathrm{RLL}(\epsilon, \sigma) = \frac{\sum_A \NLL (\epsilon (A), \sigma(A)) - \NLL (\epsilon (A), \mathrm{RMSE})}{\sum_A \NLL (\epsilon(A), |\epsilon (A)|) - \NLL (\epsilon (A), \mathrm{RMSE})}. \label{si-eq:rll}
\end{align}
NLL is the negative log-likelihood under the assumption of a Gaussian probability distribution $p(\epsilon | \sigma)$
\begin{align}
    \NLL (\epsilon, \sigma) = \frac{1}{2} \left( \frac{\epsilon^2}{\sigma^2} + \mathrm{ln} 2 \pi \sigma^2 \right).
\end{align}
RLL essentially compares the proposed uncertainty estimator $\sigma$ to an optimal estimator of the error $|\epsilon|$, by measuring how much each estimator improves a very crude estimator like the RMSE of the independent validation set.
By dividing the two values, we obtain a measure with an upper bound of one, inferring a perfect estimation.
RLL has no technical lower bound as the proposed estimator can be worse than the RMSE, resulting in a positive numerator and negative denominator in equation (\ref{si-eq:rll}).

Figure~\ref{si-fig:cc-3bpa} shows the correlation coefficients for energies and forces of the committees trained on the 3BPA dataset~\cite{Kovacs2021/10.1021/acs.jctc.1c00647} on two different test sets.
The first test set consists of structures from an MD simulation at 300\,K, which is the same ensemble as the training data.
The second test set is the same as the one used in the main article, which contains structures from a trajectory at 1200\,K.
The correlation for the energy predictions shown in the top panel is close to 0 for the 300\,K dataset, indicating that there is no correlation between the committee disagreement and the energy error.
For the out-of-domain test set at 1200K, the correlation coefficient indicates a moderate correlation, confirming our observations in Figure~\ref{fig:3BPA} of the main article.
The bottom panel of Figure~\ref{si-fig:cc-3bpa} presents the correlation coefficients for the force predictions.
It is of similar height for both test sets and signals a strong correlation between uncertainty and error.

In Figure~\ref{si-fig:rll-cc-rmd17}, the values of the Pearson correlation coefficient and RLL are shown for the rMD17 dataset~\cite{Christensen2020/10.1088/2632-2153/abba6f}, broken into individual molecules.
Overall, the measures confirm all of our observations in the main manuscript.
The uncertainties of both multi-head committees and the naive committee correlate well with their respective error.
The performance of the disjoint training set is closer to the performance of the naive committee than that of the overlapping training set.
An interesting observation is the different performance of the two measures for benzene, the simplest molecule in the dataset.
For benzene, we measure the lowest correlation coefficients, but the highest RLLs.
We argue that this difference is because RLL is based on how much the new measure improves the uncertainty estimation compared to the RMSE of an independent validation set.
The validation set is taken from the same distribution as the training data and therefore contains an equal number of structures from every molecule.
Because benzene is the simplest molecule in the dataset, the errors for it are considerably lower than for other molecules.
Therefore, the RMSE of the entire rMD17 dataset is a particularly bad estimator for benzene, and thus, the new uncertainty measure will be a more substantial improvement than for other molecules.
This results in a higher numerator in equation (\ref{si-eq:rll}) and consequently a high RLL.
It would be possible to separate the RMSE for each molecule, but this would become very complicated for large and diverse datasets and impossible for any model predicting on unknown molecules.

\section{Hold-out test in rMD17}
\label{si-sec:missing_molecules}
\begin{figure}
\centering
\includegraphics[width=\linewidth]{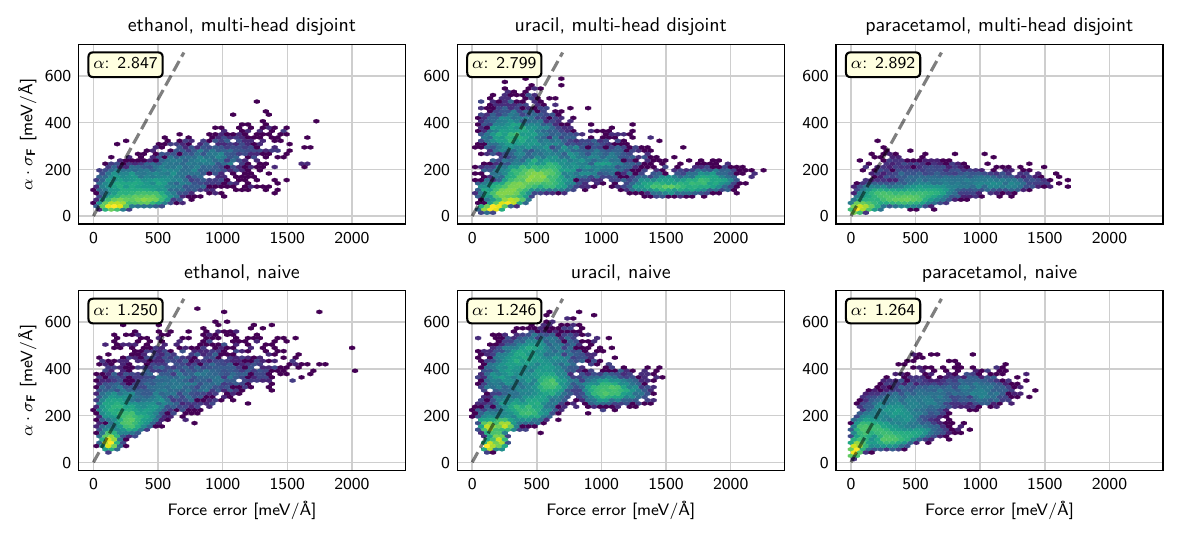}
\caption{\label{si-fig:missing-molecules}
This Figure shows the 2-dimensional distribution of the scaled force committee disagreement against force errors.
The title of each subplot shows the type of committee used, and the molecule that is missing from the training data and used for the test set.
The number in the top left of each plot indicates the scaling parameter used.
}
\end{figure}

We use the rMD17 dataset~\cite{Christensen2020/10.1088/2632-2153/abba6f} to examine how the committee disagreement performs if the investigated system is not represented in the training data.
Therefore, we exclude one of the ten molecules from the training data and train new models on the reduced dataset.
We perform this task with three different molecules: Ethanol, paracetamol, and uracil.
Afterwards, we evaluate the models on test sets consisting of structures from the molecule that was taken out of the training set.
For consistency, we scale the uncertainties with a factor calculated from the validation set comprising the same nine molecules of the training set.
Figure~\ref{si-fig:missing-molecules} shows how well the scaled committee uncertainty correlates with the true error of the model.
There are two crucial observations.
First, most data points in the distribution are to the right of the grey dashed line, which indicates perfect correlation, and, therefore, the uncertainty of most predictions is underestimated.
Secondly, there are clearly visible clusters in the distribution, most notably for the uracil test set.
Upon further investigation of these clusters, it becomes evident that each cluster can be assigned to one or multiple atoms of the molecule.
For example, in the uracil molecule, the cluster for which the uncertainty is underestimated the strongest belongs to the carbon atoms in the ring.
Since there are many molecules with carbon rings in the remaining rMD17 dataset, the local atomic environment of the carbon atoms in uracil is likely similar to that of structures present in the training data.
This leads to a biasing of the committee members' predictions, and therefore to a small committee disagreement.
However, uracil also has some unique characteristics within the rMD17 dataset.
It is the only molecule in the dataset with nitrogen atoms in the rings.
We argue that this results in a bad extrapolation and, therefore, in a high error.
We conclude that committees are susceptible to overestimating their familiarity with structures not present in the training data, resulting in an overconfident uncertainty prediction.
This will likely also play a role in foundation models, although the rich heterogeneity of the training datasets will reduce the severity.
For the same reasons, a structured analysis of this effect is unfeasible in foundation model datasets.

\section{Multi-head foundation models trained on 1000 structures}
\label{si-sec:1000_structures}
\begin{figure}
\centering
\includegraphics[width=\linewidth]{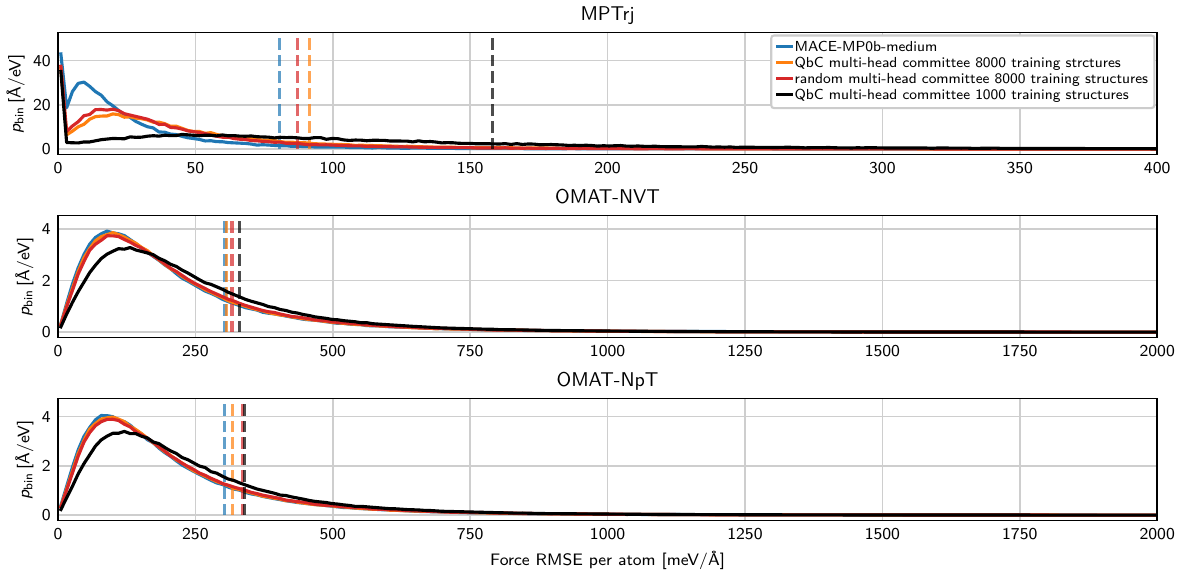}
\caption{\label{si-fig:small-foundation-committee}
The distributions of force RMSEs per atom for the MPTrj, OMAT NVT and OMAT NpT test set.
For each test set the distribution of the original MACE-MP0 foundation model, the models trained on 8000 structures (sampled using QbC and random selection) and the model trained on 1000 structures (QbC selected) is shown.
}
\end{figure}

To see how far one can condense the original MPTrj training dataset~\cite{Deng2023/10.1038/s42256-023-00716-3}, we trained a multi-head committee of the MACE-MP0 foundation~\cite{Batatia2023/10.48550/arXiv.2401.00096} model on the first 1000 structures selected in the query by committee workflow~\cite{Schran2020/10.1063/5.0016004}.
This further reduces the training dataset to 1/8 of the size of the already reduced size.
All other training parameters remained the same.
Figure~\ref{si-fig:small-foundation-committee} shows that this results in a notable decrease in performance.
While the differences between the original pre-trained head of the foundation model and the newly attached multi-head committee trained on 8000 structures in the main article were small, it becomes evident that we reached the limit of condensing the training data.
When comparing the prediction accuracy of different multi-head committees on the unknown data from the OMAT dataset~\cite{Barroso-Luque/10.48550/arXiv.2410.12771}, an interesting pattern emerges.
The distribution of errors of the two 8000 structure training sets, one selected using QbC, the other sampled randomly, is very similar.
In contrast, the 1000-structure training set performs considerably worse.
However, the RMSEs of the entire test set are very similar for all three models.
This indicates that for structures, where all three models perform poorly, the differences between the models remain small, whereas for the bulk of the test set, the larger training set leads to more accurate predictions.
Overall, the advantages of the 8000-structure condensed training set are obvious and therefore should be the preferred option.

\section{Finetuned committee}
\label{si-sec:finetuned_committee}
\begin{figure}
\centering
\includegraphics[width=\linewidth]{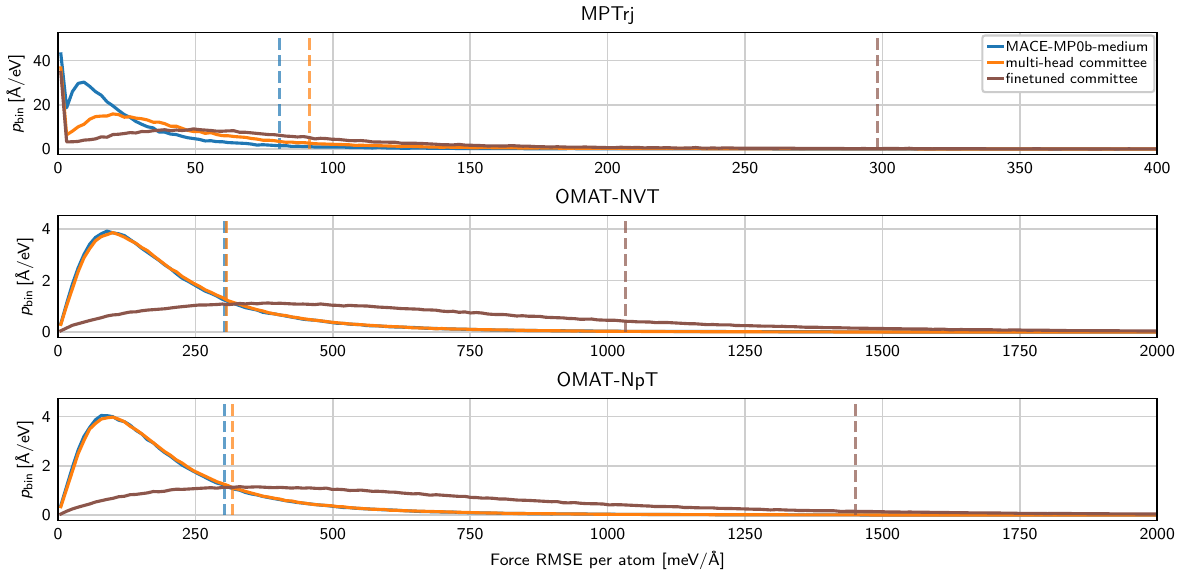}
\caption{\label{si-fig:finetuned-committee}
This figure shows the distribution of force RMSEs per atom for three different test
sets (MPtrj, OMAT NVT and OMAT NpT).
For each test set, the same three models are compared:
The original MACE-MP0 foundation model, a multi-head committee trained on 8000 QbC selected structures and a committee consisting of eight foundation models fine-tuned using the same 8000 structures.
}
\end{figure}

To compare the multi-head committee with other methods of implementing committees for foundation models, we create a naive committee of eight fine-tuned single-head foundation models~\cite{Kaur2025/10.1039/D4FD00107A}.
We use the same QbC selected training data and the same disjoint split as we used for the multi-head committee in the main article, and use 1/8 of the optimisation steps for training to keep the total computational costs constant.
Figure~\ref{si-fig:finetuned-committee} shows the distribution of force RMSE per atom, which are roughly four times worse than the errors of the original foundation models or our multi-head committee.
We kept the training procedure between the two methods as similar as possible to obtain an easy way of comparing the two workflows.
Other procedures with larger subsets of the original MPTrj dataset are likely to result in a better performance of the naive fine-tuned committee.
However, these would certainly come with a higher computational cost, would likely not outperform the multi-head committee, and would lose the benefit of obtaining a condensed training dataset.

\nolinenumbers
\section*{References}

\putbib

\end{bibunit}

\begin{thebibliography}{49}%
\makeatletter
\providecommand \@ifxundefined [1]{%
 \@ifx{#1\undefined}
}%
\providecommand \@ifnum [1]{%
 \ifnum #1\expandafter \@firstoftwo
 \else \expandafter \@secondoftwo
 \fi
}%
\providecommand \@ifx [1]{%
 \ifx #1\expandafter \@firstoftwo
 \else \expandafter \@secondoftwo
 \fi
}%
\providecommand \natexlab [1]{#1}%
\providecommand \enquote  [1]{``#1''}%
\providecommand \bibnamefont  [1]{#1}%
\providecommand \bibfnamefont [1]{#1}%
\providecommand \citenamefont [1]{#1}%
\providecommand \href@noop [0]{\@secondoftwo}%
\providecommand \href [0]{\begingroup \@sanitize@url \@href}%
\providecommand \@href[1]{\@@startlink{#1}\@@href}%
\providecommand \@@href[1]{\endgroup#1\@@endlink}%
\providecommand \@sanitize@url [0]{\catcode `\\12\catcode `\$12\catcode
  `\&12\catcode `\#12\catcode `\^12\catcode `\_12\catcode `\%12\relax}%
\providecommand \@@startlink[1]{}%
\providecommand \@@endlink[0]{}%
\providecommand \url  [0]{\begingroup\@sanitize@url \@url }%
\providecommand \@url [1]{\endgroup\@href {#1}{\urlprefix }}%
\providecommand \urlprefix  [0]{URL }%
\providecommand \Eprint [0]{\href }%
\providecommand \doibase [0]{https://doi.org/}%
\providecommand \selectlanguage [0]{\@gobble}%
\providecommand \bibinfo  [0]{\@secondoftwo}%
\providecommand \bibfield  [0]{\@secondoftwo}%
\providecommand \translation [1]{[#1]}%
\providecommand \BibitemOpen [0]{}%
\providecommand \bibitemStop [0]{}%
\providecommand \bibitemNoStop [0]{.\EOS\space}%
\providecommand \EOS [0]{\spacefactor3000\relax}%
\providecommand \BibitemShut  [1]{\csname bibitem#1\endcsname}%
\let\auto@bib@innerbib\@empty
\bibitem [{\citenamefont
  {Behler}(2021)}]{Behler2021/10.1021/acs.chemrev.0c00868}%
  \BibitemOpen
  \bibfield  {author} {\bibinfo {author} {\bibfnamefont {Jörg}\ \bibnamefont
  {Behler}},\ }\bibfield  {title} {\enquote {\bibinfo {title} {Four generations
  of high-dimensional neural network potentials},}\ }\href
  {https://doi.org/10.1021/acs.chemrev.0c00868} {\bibfield  {journal} {\bibinfo
   {journal} {Chem. Rev.}\ }\textbf {\bibinfo {volume} {121}},\ \bibinfo
  {pages} {10037--10072} (\bibinfo {year} {2021})}\BibitemShut {NoStop}%
\bibitem [{\citenamefont {Martin-Barrios}\ \emph {et~al.}(2024)\citenamefont
  {Martin-Barrios}, \citenamefont {Navas-Conyedo}, \citenamefont {Zhang},
  \citenamefont {Chen},\ and\ \citenamefont
  {Gulín-González}}]{MartinBarrios2024/10.1002/qua.27389}%
  \BibitemOpen
  \bibfield  {author} {\bibinfo {author} {\bibfnamefont {Raidel}\ \bibnamefont
  {Martin-Barrios}}, \bibinfo {author} {\bibfnamefont {Edisel}\ \bibnamefont
  {Navas-Conyedo}}, \bibinfo {author} {\bibfnamefont {Xuyi}\ \bibnamefont
  {Zhang}}, \bibinfo {author} {\bibfnamefont {Yunwei}\ \bibnamefont {Chen}},\
  and\ \bibinfo {author} {\bibfnamefont {Jorge}\ \bibnamefont
  {Gulín-González}},\ }\bibfield  {title} {\enquote {\bibinfo {title} {An
  overview about neural networks potentials in molecular dynamics
  simulation},}\ }\href {https://doi.org/https://doi.org/10.1002/qua.27389}
  {\bibfield  {journal} {\bibinfo  {journal} {Int. J. Quantum Chem.}\ }\textbf
  {\bibinfo {volume} {124}},\ \bibinfo {pages} {e27389} (\bibinfo {year}
  {2024})}\BibitemShut {NoStop}%
\bibitem [{\citenamefont {Bartók}\ \emph {et~al.}(2010)\citenamefont
  {Bartók}, \citenamefont {Payne}, \citenamefont {Kondor},\ and\ \citenamefont
  {Csányi}}]{Bartok2010/10.1103/PhysRevLett.104.136403}%
  \BibitemOpen
  \bibfield  {author} {\bibinfo {author} {\bibfnamefont {Albert~P.}\
  \bibnamefont {Bartók}}, \bibinfo {author} {\bibfnamefont {Mike~C.}\
  \bibnamefont {Payne}}, \bibinfo {author} {\bibfnamefont {Risi}\ \bibnamefont
  {Kondor}},\ and\ \bibinfo {author} {\bibfnamefont {Gábor}\ \bibnamefont
  {Csányi}},\ }\bibfield  {title} {\enquote {\bibinfo {title} {Gaussian
  approximation potentials: The accuracy of quantum mechanics, without the
  electrons},}\ }\href {https://doi.org/10.1103/PhysRevLett.104.136403}
  {\bibfield  {journal} {\bibinfo  {journal} {Phys. Rev. Lett.}\ }\textbf
  {\bibinfo {volume} {104}},\ \bibinfo {pages} {136403} (\bibinfo {year}
  {2010})}\BibitemShut {NoStop}%
\bibitem [{\citenamefont {Behler}\ and\ \citenamefont
  {Parrinello}(2007)}]{Behler2007/10.1103/PhysRevLett.98.146401}%
  \BibitemOpen
  \bibfield  {author} {\bibinfo {author} {\bibfnamefont {Jörg}\ \bibnamefont
  {Behler}}\ and\ \bibinfo {author} {\bibfnamefont {Michele}\ \bibnamefont
  {Parrinello}},\ }\bibfield  {title} {\enquote {\bibinfo {title} {Generalized
  neural-network representation of high-dimensional potential-energy
  surfaces},}\ }\href {https://doi.org/10.1103/PhysRevLett.98.146401}
  {\bibfield  {journal} {\bibinfo  {journal} {Phys. Rev. Lett.}\ }\textbf
  {\bibinfo {volume} {98}},\ \bibinfo {pages} {146401} (\bibinfo {year}
  {2007})}\BibitemShut {NoStop}%
\bibitem [{\citenamefont {Behler}(2016)}]{Behler2016/10.1063/1.4966192}%
  \BibitemOpen
  \bibfield  {author} {\bibinfo {author} {\bibfnamefont {Jörg}\ \bibnamefont
  {Behler}},\ }\bibfield  {title} {\enquote {\bibinfo {title} {Perspective:
  Machine learning potentials for atomistic simulations},}\ }\href
  {https://doi.org/10.1063/1.4966192} {\bibfield  {journal} {\bibinfo
  {journal} {J. Chem. Phys.}\ }\textbf {\bibinfo {volume} {145}} (\bibinfo
  {year} {2016}),\ 10.1063/1.4966192}\BibitemShut {NoStop}%
\bibitem [{\citenamefont {Gastegger}, \citenamefont {Behler},\ and\
  \citenamefont {Marquetand}(2017)}]{Gastegger2017/10.1039/C7SC02267K}%
  \BibitemOpen
  \bibfield  {author} {\bibinfo {author} {\bibfnamefont {Michael}\ \bibnamefont
  {Gastegger}}, \bibinfo {author} {\bibfnamefont {Jörg}\ \bibnamefont
  {Behler}},\ and\ \bibinfo {author} {\bibfnamefont {Philipp}\ \bibnamefont
  {Marquetand}},\ }\bibfield  {title} {\enquote {\bibinfo {title} {Machine
  learning molecular dynamics for the simulation of infrared spectra},}\ }\href
  {https://doi.org/10.1039/C7SC02267K} {\bibfield  {journal} {\bibinfo
  {journal} {Chem. Sci.}\ }\textbf {\bibinfo {volume} {8}},\ \bibinfo {pages}
  {6924--6935} (\bibinfo {year} {2017})}\BibitemShut {NoStop}%
\bibitem [{\citenamefont {Unke}\ \emph {et~al.}(2021)\citenamefont {Unke},
  \citenamefont {Chmiela}, \citenamefont {Sauceda}, \citenamefont {Gastegger},
  \citenamefont {Poltavsky}, \citenamefont {Schütt}, \citenamefont
  {Tkatchenko},\ and\ \citenamefont
  {Müller}}]{Unke2021/10.1021/acs.chemrev.0c01111}%
  \BibitemOpen
  \bibfield  {author} {\bibinfo {author} {\bibfnamefont {Oliver~T}\
  \bibnamefont {Unke}}, \bibinfo {author} {\bibfnamefont {Stefan}\ \bibnamefont
  {Chmiela}}, \bibinfo {author} {\bibfnamefont {Huziel~E}\ \bibnamefont
  {Sauceda}}, \bibinfo {author} {\bibfnamefont {Michael}\ \bibnamefont
  {Gastegger}}, \bibinfo {author} {\bibfnamefont {Igor}\ \bibnamefont
  {Poltavsky}}, \bibinfo {author} {\bibfnamefont {Kristof~T}\ \bibnamefont
  {Schütt}}, \bibinfo {author} {\bibfnamefont {Alexandre}\ \bibnamefont
  {Tkatchenko}},\ and\ \bibinfo {author} {\bibfnamefont {Klaus-Robert}\
  \bibnamefont {Müller}},\ }\bibfield  {title} {\enquote {\bibinfo {title}
  {Machine learning force fields},}\ }\href
  {https://doi.org/10.1021/acs.chemrev.0c01111} {\bibfield  {journal} {\bibinfo
   {journal} {Chem. Rev.}\ }\textbf {\bibinfo {volume} {121}},\ \bibinfo
  {pages} {10142--10186} (\bibinfo {year} {2021})}\BibitemShut {NoStop}%
\bibitem [{\citenamefont {Mortazavi}\ \emph {et~al.}(2023)\citenamefont
  {Mortazavi}, \citenamefont {Zhuang}, \citenamefont {Rabczuk},\ and\
  \citenamefont {Shapeev}}]{Mortazavi2023/10.1039/D3MH00125C}%
  \BibitemOpen
  \bibfield  {author} {\bibinfo {author} {\bibfnamefont {Bohayra}\ \bibnamefont
  {Mortazavi}}, \bibinfo {author} {\bibfnamefont {Xiaoying}\ \bibnamefont
  {Zhuang}}, \bibinfo {author} {\bibfnamefont {Timon}\ \bibnamefont
  {Rabczuk}},\ and\ \bibinfo {author} {\bibfnamefont {Alexander~V}\
  \bibnamefont {Shapeev}},\ }\bibfield  {title} {\enquote {\bibinfo {title}
  {Atomistic modeling of the mechanical properties: the rise of machine
  learning interatomic potentials},}\ }\href
  {https://doi.org/10.1039/D3MH00125C} {\bibfield  {journal} {\bibinfo
  {journal} {Mater. Horiz.}\ }\textbf {\bibinfo {volume} {10}},\ \bibinfo
  {pages} {1956--1968} (\bibinfo {year} {2023})}\BibitemShut {NoStop}%
\bibitem [{\citenamefont {Sch\"{u}tt}\ \emph {et~al.}(2017)\citenamefont
  {Sch\"{u}tt}, \citenamefont {Kindermans}, \citenamefont {Sauceda~Felix},
  \citenamefont {Chmiela}, \citenamefont {Tkatchenko},\ and\ \citenamefont
  {M\"{u}ller}}]{Schutt2017/AiNIP30}%
  \BibitemOpen
  \bibfield  {author} {\bibinfo {author} {\bibfnamefont {Kristof}\ \bibnamefont
  {Sch\"{u}tt}}, \bibinfo {author} {\bibfnamefont {Pieter-Jan}\ \bibnamefont
  {Kindermans}}, \bibinfo {author} {\bibfnamefont {Huziel~Enoc}\ \bibnamefont
  {Sauceda~Felix}}, \bibinfo {author} {\bibfnamefont {Stefan}\ \bibnamefont
  {Chmiela}}, \bibinfo {author} {\bibfnamefont {Alexandre}\ \bibnamefont
  {Tkatchenko}},\ and\ \bibinfo {author} {\bibfnamefont {Klaus-Robert}\
  \bibnamefont {M\"{u}ller}},\ }\bibfield  {title} {\enquote {\bibinfo {title}
  {{SchNet}: A continuous-filter convolutional neural network for modeling
  quantum interactions},}\ }in\ \href
  {https://proceedings.neurips.cc/paper_files/paper/2017/file/303ed4c69846ab36c2904d3ba8573050-Paper.pdf}
  {\emph {\bibinfo {booktitle} {Advances in Neural Information Processing
  Systems}}},\ Vol.~\bibinfo {volume} {30}\ (\bibinfo {year} {2017})\ pp.\
  \bibinfo {pages} {991--1001}\BibitemShut {NoStop}%
\bibitem [{\citenamefont
  {Drautz}(2019)}]{Drautz2019/10.1103/PhysRevB.99.014104}%
  \BibitemOpen
  \bibfield  {author} {\bibinfo {author} {\bibfnamefont {Ralf}\ \bibnamefont
  {Drautz}},\ }\bibfield  {title} {\enquote {\bibinfo {title} {Atomic cluster
  expansion for accurate and transferable interatomic potentials},}\ }\href
  {https://doi.org/10.1103/PhysRevB.99.014104} {\bibfield  {journal} {\bibinfo
  {journal} {Phys. Rev. B}\ }\textbf {\bibinfo {volume} {99}} (\bibinfo {year}
  {2019}),\ 10.1103/PhysRevB.99.014104}\BibitemShut {NoStop}%
\bibitem [{\citenamefont {Kondor}\ and\ \citenamefont
  {Trivedi}(2018)}]{Kondor2018/10.48550/arXiv.1802.03690}%
  \BibitemOpen
  \bibfield  {author} {\bibinfo {author} {\bibfnamefont {Risi}\ \bibnamefont
  {Kondor}}\ and\ \bibinfo {author} {\bibfnamefont {Shubhendu}\ \bibnamefont
  {Trivedi}},\ }\bibfield  {title} {\enquote {\bibinfo {title} {On the
  generalization of equivariance and convolution in neural networks to the
  action of compact groups},}\ }\href
  {https://doi.org/10.48550/arXiv.1802.03690} {\bibfield  {journal} {\bibinfo
  {journal} {International conference on machine learning}\ }\textbf {\bibinfo
  {volume} {80}},\ \bibinfo {pages} {2747--2755} (\bibinfo {year}
  {2018})}\BibitemShut {NoStop}%
\bibitem [{\citenamefont {Thomas}\ \emph {et~al.}(2018)\citenamefont {Thomas},
  \citenamefont {Smidt}, \citenamefont {Kearnes}, \citenamefont {Yang},
  \citenamefont {Li}, \citenamefont {Kohlhoff},\ and\ \citenamefont
  {Riley}}]{Thomas2018/10.48550/arXiv.1802.08219}%
  \BibitemOpen
  \bibfield  {author} {\bibinfo {author} {\bibfnamefont {Nathaniel}\
  \bibnamefont {Thomas}}, \bibinfo {author} {\bibfnamefont {Tess}\ \bibnamefont
  {Smidt}}, \bibinfo {author} {\bibfnamefont {Steven}\ \bibnamefont {Kearnes}},
  \bibinfo {author} {\bibfnamefont {Lusann}\ \bibnamefont {Yang}}, \bibinfo
  {author} {\bibfnamefont {Li}~\bibnamefont {Li}}, \bibinfo {author}
  {\bibfnamefont {Kai}\ \bibnamefont {Kohlhoff}},\ and\ \bibinfo {author}
  {\bibfnamefont {Patrick}\ \bibnamefont {Riley}},\ }\bibfield  {title}
  {\enquote {\bibinfo {title} {Tensor field networks: Rotation- and
  translation-equivariant neural networks for {3D} point clouds},}\ }\href
  {https://doi.org/10.48550/arXiv.1802.08219} {\bibfield  {journal} {\bibinfo
  {journal} {arxiv preprint}\ } (\bibinfo {year} {2018}),\
  10.48550/arXiv.1802.08219},\ \Eprint {https://arxiv.org/abs/1802.08219}
  {1802.08219} \BibitemShut {NoStop}%
\bibitem [{\citenamefont {Geiger}\ and\ \citenamefont
  {Smidt}(2022)}]{Geiger2022/10.48550/arXiv.2207.09453}%
  \BibitemOpen
  \bibfield  {author} {\bibinfo {author} {\bibfnamefont {Mario}\ \bibnamefont
  {Geiger}}\ and\ \bibinfo {author} {\bibfnamefont {Tess}\ \bibnamefont
  {Smidt}},\ }\bibfield  {title} {\enquote {\bibinfo {title} {e3nn: Euclidean
  neural networks},}\ }\href {https://doi.org/10.48550/arXiv.2207.09453}
  {\bibfield  {journal} {\bibinfo  {journal} {arxiv preprint}\ } (\bibinfo
  {year} {2022}),\ 10.48550/arXiv.2207.09453},\ \Eprint
  {https://arxiv.org/abs/2207.09453} {2207.09453} \BibitemShut {NoStop}%
\bibitem [{\citenamefont {Batzner}\ \emph {et~al.}(2022)\citenamefont
  {Batzner}, \citenamefont {Musaelian}, \citenamefont {Sun}, \citenamefont
  {Geiger}, \citenamefont {Mailoa}, \citenamefont {Kornbluth}, \citenamefont
  {Molinari}, \citenamefont {Smidt},\ and\ \citenamefont
  {Kozinsky}}]{Batzner2022/10.1038/s41467-022-29939-5}%
  \BibitemOpen
  \bibfield  {author} {\bibinfo {author} {\bibfnamefont {Simon}\ \bibnamefont
  {Batzner}}, \bibinfo {author} {\bibfnamefont {Albert}\ \bibnamefont
  {Musaelian}}, \bibinfo {author} {\bibfnamefont {Lixin}\ \bibnamefont {Sun}},
  \bibinfo {author} {\bibfnamefont {Mario}\ \bibnamefont {Geiger}}, \bibinfo
  {author} {\bibfnamefont {Jonathan~P.}\ \bibnamefont {Mailoa}}, \bibinfo
  {author} {\bibfnamefont {Mordechai}\ \bibnamefont {Kornbluth}}, \bibinfo
  {author} {\bibfnamefont {Nicola}\ \bibnamefont {Molinari}}, \bibinfo {author}
  {\bibfnamefont {Tess~E.}\ \bibnamefont {Smidt}},\ and\ \bibinfo {author}
  {\bibfnamefont {Boris}\ \bibnamefont {Kozinsky}},\ }\bibfield  {title}
  {\enquote {\bibinfo {title} {{E(3)}-equivariant graph neural networks for
  data-efficient and accurate interatomic potentials},}\ }\href
  {https://doi.org/10.1038/s41467-022-29939-5} {\bibfield  {journal} {\bibinfo
  {journal} {Nat. Commun.}\ }\textbf {\bibinfo {volume} {13}},\ \bibinfo
  {pages} {2453} (\bibinfo {year} {2022})}\BibitemShut {NoStop}%
\bibitem [{\citenamefont {Musaelian}\ \emph {et~al.}(2023)\citenamefont
  {Musaelian}, \citenamefont {Batzner}, \citenamefont {Johansson},
  \citenamefont {Sun}, \citenamefont {Owen}, \citenamefont {Kornbluth},\ and\
  \citenamefont {Kozinsky}}]{Musaelian2023/10.1038/s41467-023-36329-y}%
  \BibitemOpen
  \bibfield  {author} {\bibinfo {author} {\bibfnamefont {Albert}\ \bibnamefont
  {Musaelian}}, \bibinfo {author} {\bibfnamefont {Simon}\ \bibnamefont
  {Batzner}}, \bibinfo {author} {\bibfnamefont {Anders}\ \bibnamefont
  {Johansson}}, \bibinfo {author} {\bibfnamefont {Lixin}\ \bibnamefont {Sun}},
  \bibinfo {author} {\bibfnamefont {Cameron~J.}\ \bibnamefont {Owen}}, \bibinfo
  {author} {\bibfnamefont {Mordechai}\ \bibnamefont {Kornbluth}},\ and\
  \bibinfo {author} {\bibfnamefont {Boris}\ \bibnamefont {Kozinsky}},\
  }\bibfield  {title} {\enquote {\bibinfo {title} {Learning local equivariant
  representations for large-scale atomistic dynamics},}\ }\href
  {https://doi.org/10.1038/s41467-023-36329-y} {\bibfield  {journal} {\bibinfo
  {journal} {Nat. Commun.}\ }\textbf {\bibinfo {volume} {14}},\ \bibinfo
  {pages} {579} (\bibinfo {year} {2023})}\BibitemShut {NoStop}%
\bibitem [{\citenamefont {Batatia}\ \emph {et~al.}(2022)\citenamefont
  {Batatia}, \citenamefont {Kovacs}, \citenamefont {Simm}, \citenamefont
  {Ortner},\ and\ \citenamefont {Csanyi}}]{Batatia2022/MACE}%
  \BibitemOpen
  \bibfield  {author} {\bibinfo {author} {\bibfnamefont {Ilyes}\ \bibnamefont
  {Batatia}}, \bibinfo {author} {\bibfnamefont {David~P}\ \bibnamefont
  {Kovacs}}, \bibinfo {author} {\bibfnamefont {Gregor}\ \bibnamefont {Simm}},
  \bibinfo {author} {\bibfnamefont {Christoph}\ \bibnamefont {Ortner}},\ and\
  \bibinfo {author} {\bibfnamefont {Gabor}\ \bibnamefont {Csanyi}},\ }\bibfield
   {title} {\enquote {\bibinfo {title} {{MACE}: Higher order equivariant
  message passing neural networks for fast and accurate force fields},}\ }in\
  \href
  {https://proceedings.neurips.cc/paper_files/paper/2022/file/4a36c3c51af11ed9f34615b81edb5bbc-Paper-Conference.pdf}
  {\emph {\bibinfo {booktitle} {Advances in Neural Information Processing
  Systems}}},\ Vol.~\bibinfo {volume} {35}\ (\bibinfo {year} {2022})\ pp.\
  \bibinfo {pages} {11423--11436}\BibitemShut {NoStop}%
\bibitem [{\citenamefont {Takamoto}, \citenamefont {Izumi},\ and\ \citenamefont
  {Li}(2022)}]{Takamoto2022/10.1016/j.commatsci.2022.111280}%
  \BibitemOpen
  \bibfield  {author} {\bibinfo {author} {\bibfnamefont {So}~\bibnamefont
  {Takamoto}}, \bibinfo {author} {\bibfnamefont {Satoshi}\ \bibnamefont
  {Izumi}},\ and\ \bibinfo {author} {\bibfnamefont {Ju}~\bibnamefont {Li}},\
  }\bibfield  {title} {\enquote {\bibinfo {title} {{TeaNet}: Universal neural
  network interatomic potential inspired by iterative electronic
  relaxations},}\ }\href {https://doi.org/10.1016/j.commatsci.2022.111280}
  {\bibfield  {journal} {\bibinfo  {journal} {Comput. Mater. Sci.}\ }\textbf
  {\bibinfo {volume} {207}} (\bibinfo {year} {2022}),\
  10.1016/j.commatsci.2022.111280}\BibitemShut {NoStop}%
\bibitem [{\citenamefont {Yin}\ \emph {et~al.}(2025)\citenamefont {Yin},
  \citenamefont {Wang}, \citenamefont {Du}, \citenamefont {Wang}, \citenamefont
  {Ying}, \citenamefont {Jia}, \citenamefont {Zhang}, \citenamefont {Du},
  \citenamefont {Gomes}, \citenamefont {Duan}, \citenamefont {Henkelman},\ and\
  \citenamefont {Xiao}}]{Yin2025/10.48550/arXiv.2501.07155}%
  \BibitemOpen
  \bibfield  {author} {\bibinfo {author} {\bibfnamefont {Bangchen}\
  \bibnamefont {Yin}}, \bibinfo {author} {\bibfnamefont {Jiaao}\ \bibnamefont
  {Wang}}, \bibinfo {author} {\bibfnamefont {Weitao}\ \bibnamefont {Du}},
  \bibinfo {author} {\bibfnamefont {Pengbo}\ \bibnamefont {Wang}}, \bibinfo
  {author} {\bibfnamefont {Penghua}\ \bibnamefont {Ying}}, \bibinfo {author}
  {\bibfnamefont {Haojun}\ \bibnamefont {Jia}}, \bibinfo {author}
  {\bibfnamefont {Zisheng}\ \bibnamefont {Zhang}}, \bibinfo {author}
  {\bibfnamefont {Yuanqi}\ \bibnamefont {Du}}, \bibinfo {author} {\bibfnamefont
  {Carla~P.}\ \bibnamefont {Gomes}}, \bibinfo {author} {\bibfnamefont {Chenru}\
  \bibnamefont {Duan}}, \bibinfo {author} {\bibfnamefont {Graeme}\ \bibnamefont
  {Henkelman}},\ and\ \bibinfo {author} {\bibfnamefont {Hai}\ \bibnamefont
  {Xiao}},\ }\bibfield  {title} {\enquote {\bibinfo {title} {{AlphaNet}:
  Scaling up local-frame-based atomistic interatomic potential},}\ }\href
  {https://doi.org/10.48550/arXiv.2501.07155} {\bibfield  {journal} {\bibinfo
  {journal} {arxiv preprint}\ } (\bibinfo {year} {2025}),\
  10.48550/arXiv.2501.07155},\ \Eprint {https://arxiv.org/abs/2501.07155}
  {2501.07155} \BibitemShut {NoStop}%
\bibitem [{\citenamefont {Leimeroth}\ \emph {et~al.}(2025)\citenamefont
  {Leimeroth}, \citenamefont {Erhard}, \citenamefont {Albe},\ and\
  \citenamefont {Rohrer}}]{Leimeroth2025/10.48550/arXiv.2505.02503}%
  \BibitemOpen
  \bibfield  {author} {\bibinfo {author} {\bibfnamefont {Niklas}\ \bibnamefont
  {Leimeroth}}, \bibinfo {author} {\bibfnamefont {Linus~C.}\ \bibnamefont
  {Erhard}}, \bibinfo {author} {\bibfnamefont {Karsten}\ \bibnamefont {Albe}},\
  and\ \bibinfo {author} {\bibfnamefont {Jochen}\ \bibnamefont {Rohrer}},\
  }\bibfield  {title} {\enquote {\bibinfo {title} {Machine-learning interatomic
  potentials from a users perspective: A comparison of accuracy, speed and data
  efficiency},}\ }\href {https://doi.org/10.48550/arXiv.2505.02503} {\bibfield
  {journal} {\bibinfo  {journal} {arxiv preprint}\ } (\bibinfo {year} {2025}),\
  10.48550/arXiv.2505.02503},\ \Eprint {https://arxiv.org/abs/2505.02503}
  {2505.02503} \BibitemShut {NoStop}%
\bibitem [{\citenamefont {Chen}\ and\ \citenamefont
  {Ong}(2022)}]{Chen2022/10.1038/s43588-022-00349-3}%
  \BibitemOpen
  \bibfield  {author} {\bibinfo {author} {\bibfnamefont {Chi}\ \bibnamefont
  {Chen}}\ and\ \bibinfo {author} {\bibfnamefont {Shyue~Ping}\ \bibnamefont
  {Ong}},\ }\bibfield  {title} {\enquote {\bibinfo {title} {A universal graph
  deep learning interatomic potential for the periodic table},}\ }\href
  {https://doi.org/10.1038/s43588-022-00349-3} {\bibfield  {journal} {\bibinfo
  {journal} {Nat. Comput. Sci.}\ }\textbf {\bibinfo {volume} {2}},\ \bibinfo
  {pages} {718--728} (\bibinfo {year} {2022})}\BibitemShut {NoStop}%
\bibitem [{\citenamefont {Deng}\ \emph {et~al.}(2023)\citenamefont {Deng},
  \citenamefont {Zhong}, \citenamefont {Jun}, \citenamefont {Riebesell},
  \citenamefont {Han}, \citenamefont {Bartel},\ and\ \citenamefont
  {Ceder}}]{Deng2023/10.1038/s42256-023-00716-3}%
  \BibitemOpen
  \bibfield  {author} {\bibinfo {author} {\bibfnamefont {Bowen}\ \bibnamefont
  {Deng}}, \bibinfo {author} {\bibfnamefont {Peichen}\ \bibnamefont {Zhong}},
  \bibinfo {author} {\bibfnamefont {Kyu~Jung}\ \bibnamefont {Jun}}, \bibinfo
  {author} {\bibfnamefont {Janosh}\ \bibnamefont {Riebesell}}, \bibinfo
  {author} {\bibfnamefont {Kevin}\ \bibnamefont {Han}}, \bibinfo {author}
  {\bibfnamefont {Christopher~J.}\ \bibnamefont {Bartel}},\ and\ \bibinfo
  {author} {\bibfnamefont {Gerbrand}\ \bibnamefont {Ceder}},\ }\bibfield
  {title} {\enquote {\bibinfo {title} {{CHGNet} as a pretrained universal
  neural network potential for charge-informed atomistic modelling},}\ }\href
  {https://doi.org/10.1038/s42256-023-00716-3} {\bibfield  {journal} {\bibinfo
  {journal} {Nat. Mach. Intell.}\ }\textbf {\bibinfo {volume} {5}},\ \bibinfo
  {pages} {1031--1041} (\bibinfo {year} {2023})}\BibitemShut {NoStop}%
\bibitem [{\citenamefont {Batatia}\ \emph {et~al.}(2023)\citenamefont
  {Batatia}, \citenamefont {Benner}, \citenamefont {Chiang}, \citenamefont
  {Elena}, \citenamefont {Kovács}, \citenamefont {Riebesell}, \citenamefont
  {Advincula}, \citenamefont {Asta}, \citenamefont {Avaylon}, \citenamefont
  {Baldwin}, \citenamefont {Berger}, \citenamefont {Bernstein}, \citenamefont
  {Bhowmik}, \citenamefont {Blau}, \citenamefont {Cărare}, \citenamefont
  {Darby}, \citenamefont {De}, \citenamefont {Pia}, \citenamefont {Deringer},
  \citenamefont {Elijošius}, \citenamefont {El-Machachi}, \citenamefont
  {Falcioni}, \citenamefont {Fako}, \citenamefont {Ferrari}, \citenamefont
  {Genreith-Schriever}, \citenamefont {George}, \citenamefont {Goodall},
  \citenamefont {Grey}, \citenamefont {Grigorev}, \citenamefont {Han},
  \citenamefont {Handley}, \citenamefont {Heenen}, \citenamefont {Hermansson},
  \citenamefont {Holm}, \citenamefont {Jaafar}, \citenamefont {Hofmann},
  \citenamefont {Jakob}, \citenamefont {Jung}, \citenamefont {Kapil},
  \citenamefont {Kaplan}, \citenamefont {Karimitari}, \citenamefont {Kermode},
  \citenamefont {Kroupa}, \citenamefont {Kullgren}, \citenamefont {Kuner},
  \citenamefont {Kuryla}, \citenamefont {Liepuoniute}, \citenamefont {Margraf},
  \citenamefont {Magdău}, \citenamefont {Michaelides}, \citenamefont {Moore},
  \citenamefont {Naik}, \citenamefont {Niblett}, \citenamefont {Norwood},
  \citenamefont {O'Neill}, \citenamefont {Ortner}, \citenamefont {Persson},
  \citenamefont {Reuter}, \citenamefont {Rosen}, \citenamefont {Schaaf},
  \citenamefont {Schran}, \citenamefont {Shi}, \citenamefont {Sivonxay},
  \citenamefont {Stenczel}, \citenamefont {Svahn}, \citenamefont {Sutton},
  \citenamefont {Swinburne}, \citenamefont {Tilly}, \citenamefont {van~der
  Oord}, \citenamefont {Varga-Umbrich}, \citenamefont {Vegge}, \citenamefont
  {Vondrák}, \citenamefont {Wang}, \citenamefont {Witt}, \citenamefont
  {Zills},\ and\ \citenamefont
  {Csányi}}]{Batatia2023/10.48550/arXiv.2401.00096}%
  \BibitemOpen
  \bibfield  {author} {\bibinfo {author} {\bibfnamefont {Ilyes}\ \bibnamefont
  {Batatia}}, \bibinfo {author} {\bibfnamefont {Philipp}\ \bibnamefont
  {Benner}}, \bibinfo {author} {\bibfnamefont {Yuan}\ \bibnamefont {Chiang}},
  \bibinfo {author} {\bibfnamefont {Alin~M.}\ \bibnamefont {Elena}}, \bibinfo
  {author} {\bibfnamefont {Dávid~P.}\ \bibnamefont {Kovács}}, \bibinfo
  {author} {\bibfnamefont {Janosh}\ \bibnamefont {Riebesell}}, \bibinfo
  {author} {\bibfnamefont {Xavier~R.}\ \bibnamefont {Advincula}}, \bibinfo
  {author} {\bibfnamefont {Mark}\ \bibnamefont {Asta}}, \bibinfo {author}
  {\bibfnamefont {Matthew}\ \bibnamefont {Avaylon}}, \bibinfo {author}
  {\bibfnamefont {William~J.}\ \bibnamefont {Baldwin}}, \bibinfo {author}
  {\bibfnamefont {Fabian}\ \bibnamefont {Berger}}, \bibinfo {author}
  {\bibfnamefont {Noam}\ \bibnamefont {Bernstein}}, \bibinfo {author}
  {\bibfnamefont {Arghya}\ \bibnamefont {Bhowmik}}, \bibinfo {author}
  {\bibfnamefont {Samuel~M.}\ \bibnamefont {Blau}}, \bibinfo {author}
  {\bibfnamefont {Vlad}\ \bibnamefont {Cărare}}, \bibinfo {author}
  {\bibfnamefont {James~P.}\ \bibnamefont {Darby}}, \bibinfo {author}
  {\bibfnamefont {Sandip}\ \bibnamefont {De}}, \bibinfo {author} {\bibfnamefont
  {Flaviano~Della}\ \bibnamefont {Pia}}, \bibinfo {author} {\bibfnamefont
  {Volker~L.}\ \bibnamefont {Deringer}}, \bibinfo {author} {\bibfnamefont
  {Rokas}\ \bibnamefont {Elijošius}}, \bibinfo {author} {\bibfnamefont
  {Zakariya}\ \bibnamefont {El-Machachi}}, \bibinfo {author} {\bibfnamefont
  {Fabio}\ \bibnamefont {Falcioni}}, \bibinfo {author} {\bibfnamefont {Edvin}\
  \bibnamefont {Fako}}, \bibinfo {author} {\bibfnamefont {Andrea~C.}\
  \bibnamefont {Ferrari}}, \bibinfo {author} {\bibfnamefont {Annalena}\
  \bibnamefont {Genreith-Schriever}}, \bibinfo {author} {\bibfnamefont
  {Janine}\ \bibnamefont {George}}, \bibinfo {author} {\bibfnamefont {Rhys
  E.~A.}\ \bibnamefont {Goodall}}, \bibinfo {author} {\bibfnamefont {Clare~P.}\
  \bibnamefont {Grey}}, \bibinfo {author} {\bibfnamefont {Petr}\ \bibnamefont
  {Grigorev}}, \bibinfo {author} {\bibfnamefont {Shuang}\ \bibnamefont {Han}},
  \bibinfo {author} {\bibfnamefont {Will}\ \bibnamefont {Handley}}, \bibinfo
  {author} {\bibfnamefont {Hendrik~H.}\ \bibnamefont {Heenen}}, \bibinfo
  {author} {\bibfnamefont {Kersti}\ \bibnamefont {Hermansson}}, \bibinfo
  {author} {\bibfnamefont {Christian}\ \bibnamefont {Holm}}, \bibinfo {author}
  {\bibfnamefont {Jad}\ \bibnamefont {Jaafar}}, \bibinfo {author}
  {\bibfnamefont {Stephan}\ \bibnamefont {Hofmann}}, \bibinfo {author}
  {\bibfnamefont {Konstantin~S.}\ \bibnamefont {Jakob}}, \bibinfo {author}
  {\bibfnamefont {Hyunwook}\ \bibnamefont {Jung}}, \bibinfo {author}
  {\bibfnamefont {Venkat}\ \bibnamefont {Kapil}}, \bibinfo {author}
  {\bibfnamefont {Aaron~D.}\ \bibnamefont {Kaplan}}, \bibinfo {author}
  {\bibfnamefont {Nima}\ \bibnamefont {Karimitari}}, \bibinfo {author}
  {\bibfnamefont {James~R.}\ \bibnamefont {Kermode}}, \bibinfo {author}
  {\bibfnamefont {Namu}\ \bibnamefont {Kroupa}}, \bibinfo {author}
  {\bibfnamefont {Jolla}\ \bibnamefont {Kullgren}}, \bibinfo {author}
  {\bibfnamefont {Matthew~C.}\ \bibnamefont {Kuner}}, \bibinfo {author}
  {\bibfnamefont {Domantas}\ \bibnamefont {Kuryla}}, \bibinfo {author}
  {\bibfnamefont {Guoda}\ \bibnamefont {Liepuoniute}}, \bibinfo {author}
  {\bibfnamefont {Johannes~T.}\ \bibnamefont {Margraf}}, \bibinfo {author}
  {\bibfnamefont {Ioan-Bogdan}\ \bibnamefont {Magdău}}, \bibinfo {author}
  {\bibfnamefont {Angelos}\ \bibnamefont {Michaelides}}, \bibinfo {author}
  {\bibfnamefont {J.~Harry}\ \bibnamefont {Moore}}, \bibinfo {author}
  {\bibfnamefont {Aakash~A.}\ \bibnamefont {Naik}}, \bibinfo {author}
  {\bibfnamefont {Samuel~P.}\ \bibnamefont {Niblett}}, \bibinfo {author}
  {\bibfnamefont {Sam~Walton}\ \bibnamefont {Norwood}}, \bibinfo {author}
  {\bibfnamefont {Niamh}\ \bibnamefont {O'Neill}}, \bibinfo {author}
  {\bibfnamefont {Christoph}\ \bibnamefont {Ortner}}, \bibinfo {author}
  {\bibfnamefont {Kristin~A.}\ \bibnamefont {Persson}}, \bibinfo {author}
  {\bibfnamefont {Karsten}\ \bibnamefont {Reuter}}, \bibinfo {author}
  {\bibfnamefont {Andrew~S.}\ \bibnamefont {Rosen}}, \bibinfo {author}
  {\bibfnamefont {Lars~L.}\ \bibnamefont {Schaaf}}, \bibinfo {author}
  {\bibfnamefont {Christoph}\ \bibnamefont {Schran}}, \bibinfo {author}
  {\bibfnamefont {Benjamin~X.}\ \bibnamefont {Shi}}, \bibinfo {author}
  {\bibfnamefont {Eric}\ \bibnamefont {Sivonxay}}, \bibinfo {author}
  {\bibfnamefont {Tamás~K.}\ \bibnamefont {Stenczel}}, \bibinfo {author}
  {\bibfnamefont {Viktor}\ \bibnamefont {Svahn}}, \bibinfo {author}
  {\bibfnamefont {Christopher}\ \bibnamefont {Sutton}}, \bibinfo {author}
  {\bibfnamefont {Thomas~D.}\ \bibnamefont {Swinburne}}, \bibinfo {author}
  {\bibfnamefont {Jules}\ \bibnamefont {Tilly}}, \bibinfo {author}
  {\bibfnamefont {Cas}\ \bibnamefont {van~der Oord}}, \bibinfo {author}
  {\bibfnamefont {Eszter}\ \bibnamefont {Varga-Umbrich}}, \bibinfo {author}
  {\bibfnamefont {Tejs}\ \bibnamefont {Vegge}}, \bibinfo {author}
  {\bibfnamefont {Martin}\ \bibnamefont {Vondrák}}, \bibinfo {author}
  {\bibfnamefont {Yangshuai}\ \bibnamefont {Wang}}, \bibinfo {author}
  {\bibfnamefont {William~C.}\ \bibnamefont {Witt}}, \bibinfo {author}
  {\bibfnamefont {Fabian}\ \bibnamefont {Zills}},\ and\ \bibinfo {author}
  {\bibfnamefont {Gábor}\ \bibnamefont {Csányi}},\ }\bibfield  {title}
  {\enquote {\bibinfo {title} {A foundation model for atomistic materials
  chemistry},}\ }\href {https://doi.org/10.48550/arXiv.2401.00096} {\bibfield
  {journal} {\bibinfo  {journal} {arxiv preprint}\ } (\bibinfo {year} {2023}),\
  10.48550/arXiv.2401.00096},\ \Eprint {https://arxiv.org/abs/2401.00096}
  {2401.00096} \BibitemShut {NoStop}%
\bibitem [{\citenamefont {Allen}\ \emph {et~al.}(2024)\citenamefont {Allen},
  \citenamefont {Lubbers}, \citenamefont {Matin}, \citenamefont {Smith},
  \citenamefont {Messerly}, \citenamefont {Tretiak},\ and\ \citenamefont
  {Barros}}]{Allen2024/10.1038/s41524-024-01339-x}%
  \BibitemOpen
  \bibfield  {author} {\bibinfo {author} {\bibfnamefont {Alice E~A}\
  \bibnamefont {Allen}}, \bibinfo {author} {\bibfnamefont {Nicholas}\
  \bibnamefont {Lubbers}}, \bibinfo {author} {\bibfnamefont {Sakib}\
  \bibnamefont {Matin}}, \bibinfo {author} {\bibfnamefont {Justin}\
  \bibnamefont {Smith}}, \bibinfo {author} {\bibfnamefont {Richard}\
  \bibnamefont {Messerly}}, \bibinfo {author} {\bibfnamefont {Sergei}\
  \bibnamefont {Tretiak}},\ and\ \bibinfo {author} {\bibfnamefont {Kipton}\
  \bibnamefont {Barros}},\ }\bibfield  {title} {\enquote {\bibinfo {title}
  {Learning together: Towards foundation models for machine learning
  interatomic potentials with meta-learning},}\ }\href
  {https://doi.org/10.1038/s41524-024-01339-x} {\bibfield  {journal} {\bibinfo
  {journal} {npj Comput. Mater.}\ }\textbf {\bibinfo {volume} {10}},\ \bibinfo
  {pages} {154} (\bibinfo {year} {2024})}\BibitemShut {NoStop}%
\bibitem [{\citenamefont {Kaur}\ \emph {et~al.}(2025)\citenamefont {Kaur},
  \citenamefont {Pia}, \citenamefont {Batatia}, \citenamefont {Advincula},
  \citenamefont {Shi}, \citenamefont {Lan}, \citenamefont {Csányi},
  \citenamefont {Michaelides},\ and\ \citenamefont
  {Kapil}}]{Kaur2025/10.1039/D4FD00107A}%
  \BibitemOpen
  \bibfield  {author} {\bibinfo {author} {\bibfnamefont {Harveen}\ \bibnamefont
  {Kaur}}, \bibinfo {author} {\bibfnamefont {Flaviano~Della}\ \bibnamefont
  {Pia}}, \bibinfo {author} {\bibfnamefont {Ilyes}\ \bibnamefont {Batatia}},
  \bibinfo {author} {\bibfnamefont {Xavier~R}\ \bibnamefont {Advincula}},
  \bibinfo {author} {\bibfnamefont {Benjamin~X}\ \bibnamefont {Shi}}, \bibinfo
  {author} {\bibfnamefont {Jinggang}\ \bibnamefont {Lan}}, \bibinfo {author}
  {\bibfnamefont {Gábor}\ \bibnamefont {Csányi}}, \bibinfo {author}
  {\bibfnamefont {Angelos}\ \bibnamefont {Michaelides}},\ and\ \bibinfo
  {author} {\bibfnamefont {Venkat}\ \bibnamefont {Kapil}},\ }\bibfield  {title}
  {\enquote {\bibinfo {title} {Data-efficient fine-tuning of foundational
  models for first-principles quality sublimation enthalpies},}\ }\href
  {https://doi.org/10.1039/D4FD00107A} {\bibfield  {journal} {\bibinfo
  {journal} {Faraday Discuss.}\ }\textbf {\bibinfo {volume} {256}},\ \bibinfo
  {pages} {120--138} (\bibinfo {year} {2025})}\BibitemShut {NoStop}%
\bibitem [{\citenamefont {Merchant}\ \emph {et~al.}(2023)\citenamefont
  {Merchant}, \citenamefont {Batzner}, \citenamefont {Schoenholz},
  \citenamefont {Aykol}, \citenamefont {Cheon},\ and\ \citenamefont
  {Cubuk}}]{Merchant2023/10.1038/s41586-023-06735-9}%
  \BibitemOpen
  \bibfield  {author} {\bibinfo {author} {\bibfnamefont {Amil}\ \bibnamefont
  {Merchant}}, \bibinfo {author} {\bibfnamefont {Simon}\ \bibnamefont
  {Batzner}}, \bibinfo {author} {\bibfnamefont {Samuel~S}\ \bibnamefont
  {Schoenholz}}, \bibinfo {author} {\bibfnamefont {Muratahan}\ \bibnamefont
  {Aykol}}, \bibinfo {author} {\bibfnamefont {Gowoon}\ \bibnamefont {Cheon}},\
  and\ \bibinfo {author} {\bibfnamefont {Ekin~Dogus}\ \bibnamefont {Cubuk}},\
  }\bibfield  {title} {\enquote {\bibinfo {title} {Scaling deep learning for
  materials discovery},}\ }\href {https://doi.org/10.1038/s41586-023-06735-9}
  {\bibfield  {journal} {\bibinfo  {journal} {Nature}\ }\textbf {\bibinfo
  {volume} {624}},\ \bibinfo {pages} {80--85} (\bibinfo {year}
  {2023})}\BibitemShut {NoStop}%
\bibitem [{\citenamefont {Yang}\ \emph {et~al.}(2024)\citenamefont {Yang},
  \citenamefont {Hu}, \citenamefont {Zhou}, \citenamefont {Liu}, \citenamefont
  {Shi}, \citenamefont {Li}, \citenamefont {Li}, \citenamefont {Chen},
  \citenamefont {Chen}, \citenamefont {Zeni}, \citenamefont {Horton},
  \citenamefont {Pinsler}, \citenamefont {Fowler}, \citenamefont {Zügner},
  \citenamefont {Xie}, \citenamefont {Smith}, \citenamefont {Sun},
  \citenamefont {Wang}, \citenamefont {Kong}, \citenamefont {Liu},
  \citenamefont {Hao},\ and\ \citenamefont
  {Lu}}]{Yang2024/10.48550/arXiv.2405.04967}%
  \BibitemOpen
  \bibfield  {author} {\bibinfo {author} {\bibfnamefont {Han}\ \bibnamefont
  {Yang}}, \bibinfo {author} {\bibfnamefont {Chenxi}\ \bibnamefont {Hu}},
  \bibinfo {author} {\bibfnamefont {Yichi}\ \bibnamefont {Zhou}}, \bibinfo
  {author} {\bibfnamefont {Xixian}\ \bibnamefont {Liu}}, \bibinfo {author}
  {\bibfnamefont {Yu}~\bibnamefont {Shi}}, \bibinfo {author} {\bibfnamefont
  {Jielan}\ \bibnamefont {Li}}, \bibinfo {author} {\bibfnamefont {Guanzhi}\
  \bibnamefont {Li}}, \bibinfo {author} {\bibfnamefont {Zekun}\ \bibnamefont
  {Chen}}, \bibinfo {author} {\bibfnamefont {Shuizhou}\ \bibnamefont {Chen}},
  \bibinfo {author} {\bibfnamefont {Claudio}\ \bibnamefont {Zeni}}, \bibinfo
  {author} {\bibfnamefont {Matthew}\ \bibnamefont {Horton}}, \bibinfo {author}
  {\bibfnamefont {Robert}\ \bibnamefont {Pinsler}}, \bibinfo {author}
  {\bibfnamefont {Andrew}\ \bibnamefont {Fowler}}, \bibinfo {author}
  {\bibfnamefont {Daniel}\ \bibnamefont {Zügner}}, \bibinfo {author}
  {\bibfnamefont {Tian}\ \bibnamefont {Xie}}, \bibinfo {author} {\bibfnamefont
  {Jake}\ \bibnamefont {Smith}}, \bibinfo {author} {\bibfnamefont {Lixin}\
  \bibnamefont {Sun}}, \bibinfo {author} {\bibfnamefont {Qian}\ \bibnamefont
  {Wang}}, \bibinfo {author} {\bibfnamefont {Lingyu}\ \bibnamefont {Kong}},
  \bibinfo {author} {\bibfnamefont {Chang}\ \bibnamefont {Liu}}, \bibinfo
  {author} {\bibfnamefont {Hongxia}\ \bibnamefont {Hao}},\ and\ \bibinfo
  {author} {\bibfnamefont {Ziheng}\ \bibnamefont {Lu}},\ }\bibfield  {title}
  {\enquote {\bibinfo {title} {{MatterSim}: A deep learning atomistic model
  across elements, temperatures and pressures},}\ }\href
  {https://doi.org/10.48550/arXiv.2405.04967} {\bibfield  {journal} {\bibinfo
  {journal} {arxiv preprint}\ } (\bibinfo {year} {2024}),\
  10.48550/arXiv.2405.04967},\ \Eprint {https://arxiv.org/abs/2405.04967}
  {2405.04967} \BibitemShut {NoStop}%
\bibitem [{\citenamefont {Bochkarev}, \citenamefont {Lysogorskiy},\ and\
  \citenamefont {Drautz}(2024)}]{Bochkarev2024/10.1103/PhysRevX.14.021036}%
  \BibitemOpen
  \bibfield  {author} {\bibinfo {author} {\bibfnamefont {Anton}\ \bibnamefont
  {Bochkarev}}, \bibinfo {author} {\bibfnamefont {Yury}\ \bibnamefont
  {Lysogorskiy}},\ and\ \bibinfo {author} {\bibfnamefont {Ralf}\ \bibnamefont
  {Drautz}},\ }\bibfield  {title} {\enquote {\bibinfo {title} {Graph atomic
  cluster expansion for semilocal interactions beyond equivariant message
  passing},}\ }\href {https://doi.org/10.1103/PhysRevX.14.021036} {\bibfield
  {journal} {\bibinfo  {journal} {Phys. Rev. X}\ }\textbf {\bibinfo {volume}
  {14}},\ \bibinfo {pages} {021036} (\bibinfo {year} {2024})}\BibitemShut
  {NoStop}%
\bibitem [{\citenamefont {Kim}\ \emph {et~al.}(2024)\citenamefont {Kim},
  \citenamefont {Kim}, \citenamefont {Kim}, \citenamefont {Lee}, \citenamefont
  {Park}, \citenamefont {Kang},\ and\ \citenamefont
  {Han}}]{Kim2024/10.1021/jacs.4c14455}%
  \BibitemOpen
  \bibfield  {author} {\bibinfo {author} {\bibfnamefont {Jaesun}\ \bibnamefont
  {Kim}}, \bibinfo {author} {\bibfnamefont {Jisu}\ \bibnamefont {Kim}},
  \bibinfo {author} {\bibfnamefont {Jaehoon}\ \bibnamefont {Kim}}, \bibinfo
  {author} {\bibfnamefont {Jiho}\ \bibnamefont {Lee}}, \bibinfo {author}
  {\bibfnamefont {Yutack}\ \bibnamefont {Park}}, \bibinfo {author}
  {\bibfnamefont {Youngho}\ \bibnamefont {Kang}},\ and\ \bibinfo {author}
  {\bibfnamefont {Seungwu}\ \bibnamefont {Han}},\ }\bibfield  {title} {\enquote
  {\bibinfo {title} {Data-efficient multi-fidelity training for high-fidelity
  machine learning interatomic potentials},}\ }\href
  {https://doi.org/10.1021/jacs.4c14455} {\bibfield  {journal} {\bibinfo
  {journal} {J. Am. Chem. Soc.}\ }\textbf {\bibinfo {volume} {147}},\ \bibinfo
  {pages} {1042--1054} (\bibinfo {year} {2024})}\BibitemShut {NoStop}%
\bibitem [{\citenamefont {Fu}\ \emph {et~al.}(2025)\citenamefont {Fu},
  \citenamefont {Wood}, \citenamefont {Barroso-Luque}, \citenamefont {Levine},
  \citenamefont {Gao}, \citenamefont {Dzamba},\ and\ \citenamefont
  {Zitnick}}]{Fu2025/10.48550/arXiv.2502.12147}%
  \BibitemOpen
  \bibfield  {author} {\bibinfo {author} {\bibfnamefont {Xiang}\ \bibnamefont
  {Fu}}, \bibinfo {author} {\bibfnamefont {Brandon~M.}\ \bibnamefont {Wood}},
  \bibinfo {author} {\bibfnamefont {Luis}\ \bibnamefont {Barroso-Luque}},
  \bibinfo {author} {\bibfnamefont {Daniel~S.}\ \bibnamefont {Levine}},
  \bibinfo {author} {\bibfnamefont {Meng}\ \bibnamefont {Gao}}, \bibinfo
  {author} {\bibfnamefont {Misko}\ \bibnamefont {Dzamba}},\ and\ \bibinfo
  {author} {\bibfnamefont {C.~Lawrence}\ \bibnamefont {Zitnick}},\ }\bibfield
  {title} {\enquote {\bibinfo {title} {Learning smooth and expressive
  interatomic potentials for physical property prediction},}\ }\href
  {https://doi.org/10.48550/arXiv.2502.12147} {\bibfield  {journal} {\bibinfo
  {journal} {arxiv preprint}\ } (\bibinfo {year} {2025}),\
  10.48550/arXiv.2502.12147},\ \Eprint {https://arxiv.org/abs/2402.14147}
  {2402.14147} \BibitemShut {NoStop}%
\bibitem [{\citenamefont {Focassio}, \citenamefont {Freitas},\ and\
  \citenamefont {Schleder}(2025)}]{Focassio2025/10.1021/acsami.4c03815}%
  \BibitemOpen
  \bibfield  {author} {\bibinfo {author} {\bibfnamefont {Bruno}\ \bibnamefont
  {Focassio}}, \bibinfo {author} {\bibfnamefont {Luis Paulo~M.}\ \bibnamefont
  {Freitas}},\ and\ \bibinfo {author} {\bibfnamefont {Gabriel~R}\ \bibnamefont
  {Schleder}},\ }\bibfield  {title} {\enquote {\bibinfo {title} {Performance
  assessment of universal machine learning interatomic potentials: Challenges
  and directions for materials’ surfaces},}\ }\href
  {https://doi.org/10.1021/acsami.4c03815} {\bibfield  {journal} {\bibinfo
  {journal} {ACS Appl. Mater. Interfaces}\ }\textbf {\bibinfo {volume} {17}},\
  \bibinfo {pages} {13111--13121} (\bibinfo {year} {2025})}\BibitemShut
  {NoStop}%
\bibitem [{\citenamefont {Imbalzano}\ \emph {et~al.}(2021)\citenamefont
  {Imbalzano}, \citenamefont {Zhuang}, \citenamefont {Kapil}, \citenamefont
  {Rossi}, \citenamefont {Engel}, \citenamefont {Grasselli},\ and\
  \citenamefont {Ceriotti}}]{Imbalzano2021/10.1063/5.0036522}%
  \BibitemOpen
  \bibfield  {author} {\bibinfo {author} {\bibfnamefont {Giulio}\ \bibnamefont
  {Imbalzano}}, \bibinfo {author} {\bibfnamefont {Yongbin}\ \bibnamefont
  {Zhuang}}, \bibinfo {author} {\bibfnamefont {Venkat}\ \bibnamefont {Kapil}},
  \bibinfo {author} {\bibfnamefont {Kevin}\ \bibnamefont {Rossi}}, \bibinfo
  {author} {\bibfnamefont {Edgar~A}\ \bibnamefont {Engel}}, \bibinfo {author}
  {\bibfnamefont {Federico}\ \bibnamefont {Grasselli}},\ and\ \bibinfo {author}
  {\bibfnamefont {Michele}\ \bibnamefont {Ceriotti}},\ }\bibfield  {title}
  {\enquote {\bibinfo {title} {Uncertainty estimation for molecular dynamics
  and sampling},}\ }\href {https://doi.org/10.1063/5.0036522} {\bibfield
  {journal} {\bibinfo  {journal} {J. Chem. Phys.}\ }\textbf {\bibinfo {volume}
  {154}},\ \bibinfo {pages} {074102} (\bibinfo {year} {2021})}\BibitemShut
  {NoStop}%
\bibitem [{\citenamefont {Dai}, \citenamefont {Adhikari},\ and\ \citenamefont
  {Wen}(2025)}]{Dai2025/10.1515/revce-2024-0028}%
  \BibitemOpen
  \bibfield  {author} {\bibinfo {author} {\bibfnamefont {Jin}\ \bibnamefont
  {Dai}}, \bibinfo {author} {\bibfnamefont {Santosh}\ \bibnamefont
  {Adhikari}},\ and\ \bibinfo {author} {\bibfnamefont {Mingjian}\ \bibnamefont
  {Wen}},\ }\bibfield  {title} {\enquote {\bibinfo {title} {Uncertainty
  quantification and propagation in atomistic machine learning},}\ }\href
  {https://doi.org/10.1515/revce-2024-0028} {\bibfield  {journal} {\bibinfo
  {journal} {Rev. Chem. Eng.}\ }\textbf {\bibinfo {volume} {41}},\ \bibinfo
  {pages} {333--357} (\bibinfo {year} {2025})}\BibitemShut {NoStop}%
\bibitem [{\citenamefont {Gawlikowski}\ \emph {et~al.}(2023)\citenamefont
  {Gawlikowski}, \citenamefont {Tassi}, \citenamefont {Ali}, \citenamefont
  {Lee}, \citenamefont {Humt}, \citenamefont {Feng}, \citenamefont {Kruspe},
  \citenamefont {Triebel}, \citenamefont {Jung}, \citenamefont {Roscher},
  \citenamefont {Shahzad}, \citenamefont {Yang}, \citenamefont {Bamler},\ and\
  \citenamefont {Zhu}}]{Gawlikowski2023/10.1007/s10462-023-10562-9}%
  \BibitemOpen
  \bibfield  {author} {\bibinfo {author} {\bibfnamefont {Jakob}\ \bibnamefont
  {Gawlikowski}}, \bibinfo {author} {\bibfnamefont {Cedrique
  Rovile~Njieutcheu}\ \bibnamefont {Tassi}}, \bibinfo {author} {\bibfnamefont
  {Mohsin}\ \bibnamefont {Ali}}, \bibinfo {author} {\bibfnamefont {Jongseok}\
  \bibnamefont {Lee}}, \bibinfo {author} {\bibfnamefont {Matthias}\
  \bibnamefont {Humt}}, \bibinfo {author} {\bibfnamefont {Jianxiang}\
  \bibnamefont {Feng}}, \bibinfo {author} {\bibfnamefont {Anna}\ \bibnamefont
  {Kruspe}}, \bibinfo {author} {\bibfnamefont {Rudolph}\ \bibnamefont
  {Triebel}}, \bibinfo {author} {\bibfnamefont {Peter}\ \bibnamefont {Jung}},
  \bibinfo {author} {\bibfnamefont {Ribana}\ \bibnamefont {Roscher}}, \bibinfo
  {author} {\bibfnamefont {Muhammad}\ \bibnamefont {Shahzad}}, \bibinfo
  {author} {\bibfnamefont {Wen}\ \bibnamefont {Yang}}, \bibinfo {author}
  {\bibfnamefont {Richard}\ \bibnamefont {Bamler}},\ and\ \bibinfo {author}
  {\bibfnamefont {Xiao~Xiang}\ \bibnamefont {Zhu}},\ }\bibfield  {title}
  {\enquote {\bibinfo {title} {A survey of uncertainty in deep neural
  networks},}\ }\href {https://doi.org/10.1007/s10462-023-10562-9} {\bibfield
  {journal} {\bibinfo  {journal} {Artif. Intell. Rev.}\ }\textbf {\bibinfo
  {volume} {56}},\ \bibinfo {pages} {1513--1589} (\bibinfo {year}
  {2023})}\BibitemShut {NoStop}%
\bibitem [{\citenamefont {Bigi}\ \emph {et~al.}(2024)\citenamefont {Bigi},
  \citenamefont {Chong}, \citenamefont {Ceriotti},\ and\ \citenamefont
  {Grasselli}}]{Bigi2024/10.1088/2632-2153/ad805f}%
  \BibitemOpen
  \bibfield  {author} {\bibinfo {author} {\bibfnamefont {Filippo}\ \bibnamefont
  {Bigi}}, \bibinfo {author} {\bibfnamefont {Sanggyu}\ \bibnamefont {Chong}},
  \bibinfo {author} {\bibfnamefont {Michele}\ \bibnamefont {Ceriotti}},\ and\
  \bibinfo {author} {\bibfnamefont {Federico}\ \bibnamefont {Grasselli}},\
  }\bibfield  {title} {\enquote {\bibinfo {title} {A prediction rigidity
  formalism for low-cost uncertainties in trained neural networks},}\ }\href
  {https://doi.org/10.1088/2632-2153/ad805f} {\bibfield  {journal} {\bibinfo
  {journal} {Mach. Learn.: Sci. Technol.}\ }\textbf {\bibinfo {volume} {5}},\
  \bibinfo {pages} {045018} (\bibinfo {year} {2024})}\BibitemShut {NoStop}%
\bibitem [{\citenamefont {Bilbrey}\ \emph {et~al.}(2025)\citenamefont
  {Bilbrey}, \citenamefont {Firoz}, \citenamefont {Lee},\ and\ \citenamefont
  {Choudhury}}]{Bilbrey2025/10.1038/s41524-025-01572-y}%
  \BibitemOpen
  \bibfield  {author} {\bibinfo {author} {\bibfnamefont {Jenna~A.}\
  \bibnamefont {Bilbrey}}, \bibinfo {author} {\bibfnamefont {Jesun~S.}\
  \bibnamefont {Firoz}}, \bibinfo {author} {\bibfnamefont {Mal~Soon}\
  \bibnamefont {Lee}},\ and\ \bibinfo {author} {\bibfnamefont {Sutanay}\
  \bibnamefont {Choudhury}},\ }\bibfield  {title} {\enquote {\bibinfo {title}
  {Uncertainty quantification for neural network potential foundation
  models},}\ }\href {https://doi.org/10.1038/s41524-025-01572-y} {\bibfield
  {journal} {\bibinfo  {journal} {npj Comput. Mater.}\ }\textbf {\bibinfo
  {volume} {11}} (\bibinfo {year} {2025}),\
  10.1038/s41524-025-01572-y}\BibitemShut {NoStop}%
\bibitem [{\citenamefont {Zhu}\ \emph {et~al.}(2023)\citenamefont {Zhu},
  \citenamefont {Batzner}, \citenamefont {Musaelian},\ and\ \citenamefont
  {Kozinsky}}]{Zhu2023/10.1063/5.0136574}%
  \BibitemOpen
  \bibfield  {author} {\bibinfo {author} {\bibfnamefont {Albert}\ \bibnamefont
  {Zhu}}, \bibinfo {author} {\bibfnamefont {Simon}\ \bibnamefont {Batzner}},
  \bibinfo {author} {\bibfnamefont {Albert}\ \bibnamefont {Musaelian}},\ and\
  \bibinfo {author} {\bibfnamefont {Boris}\ \bibnamefont {Kozinsky}},\
  }\bibfield  {title} {\enquote {\bibinfo {title} {Fast uncertainty estimates
  in deep learning interatomic potentials},}\ }\href
  {https://doi.org/10.1063/5.0136574} {\bibfield  {journal} {\bibinfo
  {journal} {J. Chem. Phys.}\ }\textbf {\bibinfo {volume} {158}},\ \bibinfo
  {pages} {164111} (\bibinfo {year} {2023})}\BibitemShut {NoStop}%
\bibitem [{\citenamefont {Schwalbe-Koda}\ \emph {et~al.}(2025)\citenamefont
  {Schwalbe-Koda}, \citenamefont {Hamel}, \citenamefont {Sadigh}, \citenamefont
  {Zhou},\ and\ \citenamefont
  {Lordi}}]{Schwalbe-Koda2025/10.1038/s41467-025-59232-0}%
  \BibitemOpen
  \bibfield  {author} {\bibinfo {author} {\bibfnamefont {Daniel}\ \bibnamefont
  {Schwalbe-Koda}}, \bibinfo {author} {\bibfnamefont {Sebastien}\ \bibnamefont
  {Hamel}}, \bibinfo {author} {\bibfnamefont {Babak}\ \bibnamefont {Sadigh}},
  \bibinfo {author} {\bibfnamefont {Fei}\ \bibnamefont {Zhou}},\ and\ \bibinfo
  {author} {\bibfnamefont {Vincenzo}\ \bibnamefont {Lordi}},\ }\bibfield
  {title} {\enquote {\bibinfo {title} {Model-free estimation of completeness,
  uncertainties, and outliers in atomistic machine learning using information
  theory},}\ }\href {https://doi.org/10.1038/s41467-025-59232-0} {\bibfield
  {journal} {\bibinfo  {journal} {Nat. Commun.}\ }\textbf {\bibinfo {volume}
  {16}},\ \bibinfo {pages} {4014} (\bibinfo {year} {2025})}\BibitemShut
  {NoStop}%
\bibitem [{\citenamefont {Schran}, \citenamefont {Brezina},\ and\ \citenamefont
  {Marsalek}(2020)}]{Schran2020/10.1063/5.0016004}%
  \BibitemOpen
  \bibfield  {author} {\bibinfo {author} {\bibfnamefont {Christoph}\
  \bibnamefont {Schran}}, \bibinfo {author} {\bibfnamefont {Krystof}\
  \bibnamefont {Brezina}},\ and\ \bibinfo {author} {\bibfnamefont {Ondrej}\
  \bibnamefont {Marsalek}},\ }\bibfield  {title} {\enquote {\bibinfo {title}
  {Committee neural network potentials control generalization errors and enable
  active learning},}\ }\href {https://doi.org/10.1063/5.0016004} {\bibfield
  {journal} {\bibinfo  {journal} {J. Chem. Phys.}\ }\textbf {\bibinfo {volume}
  {153}} (\bibinfo {year} {2020}),\ 10.1063/5.0016004}\BibitemShut {NoStop}%
\bibitem [{\citenamefont {Carrete}\ \emph {et~al.}(2023)\citenamefont
  {Carrete}, \citenamefont {Montes-Campos}, \citenamefont {Wanzenböck},
  \citenamefont {Heid},\ and\ \citenamefont
  {Madsen}}]{Carrete2023/10.1063/5.0146905}%
  \BibitemOpen
  \bibfield  {author} {\bibinfo {author} {\bibfnamefont {Jesús}\ \bibnamefont
  {Carrete}}, \bibinfo {author} {\bibfnamefont {Hadrián}\ \bibnamefont
  {Montes-Campos}}, \bibinfo {author} {\bibfnamefont {Ralf}\ \bibnamefont
  {Wanzenböck}}, \bibinfo {author} {\bibfnamefont {Esther}\ \bibnamefont
  {Heid}},\ and\ \bibinfo {author} {\bibfnamefont {Georg K~H}\ \bibnamefont
  {Madsen}},\ }\bibfield  {title} {\enquote {\bibinfo {title} {Deep ensembles
  vs committees for uncertainty estimation in neural-network force fields:
  Comparison and application to active learning},}\ }\href
  {https://doi.org/10.1063/5.0146905} {\bibfield  {journal} {\bibinfo
  {journal} {J. Chem. Phys.}\ }\textbf {\bibinfo {volume} {158}},\ \bibinfo
  {pages} {204801} (\bibinfo {year} {2023})}\BibitemShut {NoStop}%
\bibitem [{\citenamefont {Kellner}\ and\ \citenamefont
  {Ceriotti}(2024)}]{Kellner2024/10.1088/2632-2153/ad594a}%
  \BibitemOpen
  \bibfield  {author} {\bibinfo {author} {\bibfnamefont {Matthias}\
  \bibnamefont {Kellner}}\ and\ \bibinfo {author} {\bibfnamefont {Michele}\
  \bibnamefont {Ceriotti}},\ }\bibfield  {title} {\enquote {\bibinfo {title}
  {Uncertainty quantification by direct propagation of shallow ensembles},}\
  }\href {https://doi.org/10.1088/2632-2153/ad594a} {\bibfield  {journal}
  {\bibinfo  {journal} {Mach. Learn.: Sci. Technol.}\ }\textbf {\bibinfo
  {volume} {5}},\ \bibinfo {pages} {035006} (\bibinfo {year}
  {2024})}\BibitemShut {NoStop}%
\bibitem [{\citenamefont {Schaaf}\ \emph {et~al.}(2023)\citenamefont {Schaaf},
  \citenamefont {Fako}, \citenamefont {De}, \citenamefont {Schäfer},\ and\
  \citenamefont {Csányi}}]{Schaaf2023/10.1038/s41524-023-01124-2}%
  \BibitemOpen
  \bibfield  {author} {\bibinfo {author} {\bibfnamefont {Lars~L}\ \bibnamefont
  {Schaaf}}, \bibinfo {author} {\bibfnamefont {Edvin}\ \bibnamefont {Fako}},
  \bibinfo {author} {\bibfnamefont {Sandip}\ \bibnamefont {De}}, \bibinfo
  {author} {\bibfnamefont {Ansgar}\ \bibnamefont {Schäfer}},\ and\ \bibinfo
  {author} {\bibfnamefont {Gábor}\ \bibnamefont {Csányi}},\ }\bibfield
  {title} {\enquote {\bibinfo {title} {Accurate energy barriers for catalytic
  reaction pathways: an automatic training protocol for machine learning force
  fields},}\ }\href {https://doi.org/10.1038/s41524-023-01124-2} {\bibfield
  {journal} {\bibinfo  {journal} {npj Comput. Mater.}\ }\textbf {\bibinfo
  {volume} {9}},\ \bibinfo {pages} {180} (\bibinfo {year} {2023})}\BibitemShut
  {NoStop}%
\bibitem [{\citenamefont {Holzmüller}\ \emph {et~al.}(2023)\citenamefont
  {Holzmüller}, \citenamefont {Zaverkin}, \citenamefont {Kästner},\ and\
  \citenamefont {Steinwart}}]{Holzmuller2023/activeLearning}%
  \BibitemOpen
  \bibfield  {author} {\bibinfo {author} {\bibfnamefont {David}\ \bibnamefont
  {Holzmüller}}, \bibinfo {author} {\bibfnamefont {Viktor}\ \bibnamefont
  {Zaverkin}}, \bibinfo {author} {\bibfnamefont {Johannes}\ \bibnamefont
  {Kästner}},\ and\ \bibinfo {author} {\bibfnamefont {Ingo}\ \bibnamefont
  {Steinwart}},\ }\bibfield  {title} {\enquote {\bibinfo {title} {A framework
  and benchmark for deep batch active learning for regression},}\ }\href
  {http://jmlr.org/papers/v24/22-0937.html} {\bibfield  {journal} {\bibinfo
  {journal} {J. Mach. Learn. Res.}\ }\textbf {\bibinfo {volume} {24}},\
  \bibinfo {pages} {1--81} (\bibinfo {year} {2023})}\BibitemShut {NoStop}%
\bibitem [{\citenamefont {Batatia}\ \emph {et~al.}(2025)\citenamefont
  {Batatia}, \citenamefont {Batzner}, \citenamefont {Kovács}, \citenamefont
  {Musaelian}, \citenamefont {Simm}, \citenamefont {Drautz}, \citenamefont
  {Ortner}, \citenamefont {Kozinsky},\ and\ \citenamefont
  {Csányi}}]{Batatia2025/10.1038/s42256-024-00956-x}%
  \BibitemOpen
  \bibfield  {author} {\bibinfo {author} {\bibfnamefont {Ilyes}\ \bibnamefont
  {Batatia}}, \bibinfo {author} {\bibfnamefont {Simon}\ \bibnamefont
  {Batzner}}, \bibinfo {author} {\bibfnamefont {Dávid~Péter}\ \bibnamefont
  {Kovács}}, \bibinfo {author} {\bibfnamefont {Albert}\ \bibnamefont
  {Musaelian}}, \bibinfo {author} {\bibfnamefont {Gregor N~C}\ \bibnamefont
  {Simm}}, \bibinfo {author} {\bibfnamefont {Ralf}\ \bibnamefont {Drautz}},
  \bibinfo {author} {\bibfnamefont {Christoph}\ \bibnamefont {Ortner}},
  \bibinfo {author} {\bibfnamefont {Boris}\ \bibnamefont {Kozinsky}},\ and\
  \bibinfo {author} {\bibfnamefont {Gábor}\ \bibnamefont {Csányi}},\
  }\bibfield  {title} {\enquote {\bibinfo {title} {The design space of
  {E(3)}-equivariant atom-centred interatomic potentials},}\ }\href
  {https://doi.org/10.1038/s42256-024-00956-x} {\bibfield  {journal} {\bibinfo
  {journal} {Nat. Mach. Intell.}\ }\textbf {\bibinfo {volume} {7}},\ \bibinfo
  {pages} {56--67} (\bibinfo {year} {2025})}\BibitemShut {NoStop}%
\bibitem [{\citenamefont {Kahle}\ and\ \citenamefont
  {Zipoli}(2022)}]{Kahle2022/10.1103/PhysRevE.105.015311}%
  \BibitemOpen
  \bibfield  {author} {\bibinfo {author} {\bibfnamefont {Leonid}\ \bibnamefont
  {Kahle}}\ and\ \bibinfo {author} {\bibfnamefont {Federico}\ \bibnamefont
  {Zipoli}},\ }\bibfield  {title} {\enquote {\bibinfo {title} {Quality of
  uncertainty estimates from neural network potential ensembles},}\ }\href
  {https://doi.org/10.1103/PhysRevE.105.015311} {\bibfield  {journal} {\bibinfo
   {journal} {Phys. Rev. E}\ }\textbf {\bibinfo {volume} {105}},\ \bibinfo
  {pages} {15311} (\bibinfo {year} {2022})}\BibitemShut {NoStop}%
\bibitem [{\citenamefont {Pearson}(1895)}]{Pearson1895/10.1098/rspl.1895.0041}%
  \BibitemOpen
  \bibfield  {author} {\bibinfo {author} {\bibfnamefont {Karl}\ \bibnamefont
  {Pearson}},\ }\bibfield  {title} {\enquote {\bibinfo {title} {Vii. note on
  regression and inheritance in the case of two parents},}\ }\href
  {https://doi.org/10.1098/rspl.1895.0041} {\bibfield  {journal} {\bibinfo
  {journal} {Proc. R. Soc. Lond.}\ }\textbf {\bibinfo {volume} {58}},\ \bibinfo
  {pages} {240--242} (\bibinfo {year} {1895})}\BibitemShut {NoStop}%
\bibitem [{\citenamefont {Kovács}\ \emph {et~al.}(2021)\citenamefont
  {Kovács}, \citenamefont {van~der Oord}, \citenamefont {Kucera},
  \citenamefont {Allen}, \citenamefont {Cole}, \citenamefont {Ortner},\ and\
  \citenamefont {Csányi}}]{Kovacs2021/10.1021/acs.jctc.1c00647}%
  \BibitemOpen
  \bibfield  {author} {\bibinfo {author} {\bibfnamefont {Dávid~Péter}\
  \bibnamefont {Kovács}}, \bibinfo {author} {\bibfnamefont {Cas}\ \bibnamefont
  {van~der Oord}}, \bibinfo {author} {\bibfnamefont {Jiri}\ \bibnamefont
  {Kucera}}, \bibinfo {author} {\bibfnamefont {Alice E~A}\ \bibnamefont
  {Allen}}, \bibinfo {author} {\bibfnamefont {Daniel~J}\ \bibnamefont {Cole}},
  \bibinfo {author} {\bibfnamefont {Christoph}\ \bibnamefont {Ortner}},\ and\
  \bibinfo {author} {\bibfnamefont {Gábor}\ \bibnamefont {Csányi}},\
  }\bibfield  {title} {\enquote {\bibinfo {title} {Linear atomic cluster
  expansion force fields for organic molecules: Beyond {RMSE}},}\ }\href
  {https://doi.org/10.1021/acs.jctc.1c00647} {\bibfield  {journal} {\bibinfo
  {journal} {J. Chem. Theory Comput.}\ }\textbf {\bibinfo {volume} {17}},\
  \bibinfo {pages} {7696--7711} (\bibinfo {year} {2021})}\BibitemShut {NoStop}%
\bibitem [{\citenamefont {Christensen}\ and\ \citenamefont {von
  Lilienfeld}(2020)}]{Christensen2020/10.1088/2632-2153/abba6f}%
  \BibitemOpen
  \bibfield  {author} {\bibinfo {author} {\bibfnamefont {Anders~S}\
  \bibnamefont {Christensen}}\ and\ \bibinfo {author} {\bibfnamefont
  {O~Anatole}\ \bibnamefont {von Lilienfeld}},\ }\bibfield  {title} {\enquote
  {\bibinfo {title} {On the role of gradients for machine learning of molecular
  energies and forces},}\ }\href {https://doi.org/10.1088/2632-2153/abba6f}
  {\bibfield  {journal} {\bibinfo  {journal} {Mach. Learn.: Sci. Technol.}\
  }\textbf {\bibinfo {volume} {1}},\ \bibinfo {pages} {045018} (\bibinfo {year}
  {2020})}\BibitemShut {NoStop}%
\bibitem [{\citenamefont {Perdew}, \citenamefont {Burke},\ and\ \citenamefont
  {Ernzerhof}(1996)}]{Perdew1996/10.1103/PhysRevLett.77.3865}%
  \BibitemOpen
  \bibfield  {author} {\bibinfo {author} {\bibfnamefont {John~P}\ \bibnamefont
  {Perdew}}, \bibinfo {author} {\bibfnamefont {Kieron}\ \bibnamefont {Burke}},\
  and\ \bibinfo {author} {\bibfnamefont {Matthias}\ \bibnamefont {Ernzerhof}},\
  }\bibfield  {title} {\enquote {\bibinfo {title} {Generalized gradient
  approximation made simple},}\ }\href
  {https://doi.org/10.1103/PhysRevLett.77.3865} {\bibfield  {journal} {\bibinfo
   {journal} {Phys. Rev. Lett.}\ }\textbf {\bibinfo {volume} {77}},\ \bibinfo
  {pages} {3865--3868} (\bibinfo {year} {1996})}\BibitemShut {NoStop}%
\bibitem [{\citenamefont {Barroso-Luque}\ \emph {et~al.}(2024)\citenamefont
  {Barroso-Luque}, \citenamefont {Muhammed}, \citenamefont {Fu}, \citenamefont
  {Brandon}, \citenamefont {Gao}, \citenamefont {Rizvi}, \citenamefont
  {Zitnick},\ and\ \citenamefont
  {Ulissi}}]{Barroso-Luque/10.48550/arXiv.2410.12771}%
  \BibitemOpen
  \bibfield  {author} {\bibinfo {author} {\bibfnamefont {Luis}\ \bibnamefont
  {Barroso-Luque}}, \bibinfo {author} {\bibfnamefont {Shuaibi}\ \bibnamefont
  {Muhammed}}, \bibinfo {author} {\bibfnamefont {Xiang}\ \bibnamefont {Fu}},
  \bibinfo {author} {\bibfnamefont {Dzamba Misko~Wood}\ \bibnamefont
  {Brandon}}, \bibinfo {author} {\bibfnamefont {Meng}\ \bibnamefont {Gao}},
  \bibinfo {author} {\bibfnamefont {Ammar}\ \bibnamefont {Rizvi}}, \bibinfo
  {author} {\bibfnamefont {C~Lawrence}\ \bibnamefont {Zitnick}},\ and\ \bibinfo
  {author} {\bibfnamefont {Zachary~W}\ \bibnamefont {Ulissi}},\ }\bibfield
  {title} {\enquote {\bibinfo {title} {Open materials 2024 ({OMat24}) inorganic
  materials dataset and models},}\ }\href
  {https://doi.org/10.48550/arXiv.2410.12771} {\bibfield  {journal} {\bibinfo
  {journal} {arXiv preprint}\ } (\bibinfo {year} {2024}),\
  10.48550/arXiv.2410.12771},\ \Eprint {https://arxiv.org/abs/2410.12771}
  {arXiv:2410.12771 [cond-mat.mtrl-sci]} \BibitemShut {NoStop}%
\end{thebibliography}%


\begin{thebibliography}{9}%
\makeatletter
\providecommand \@ifxundefined [1]{%
 \@ifx{#1\undefined}
}%
\providecommand \@ifnum [1]{%
 \ifnum #1\expandafter \@firstoftwo
 \else \expandafter \@secondoftwo
 \fi
}%
\providecommand \@ifx [1]{%
 \ifx #1\expandafter \@firstoftwo
 \else \expandafter \@secondoftwo
 \fi
}%
\providecommand \natexlab [1]{#1}%
\providecommand \enquote  [1]{``#1''}%
\providecommand \bibnamefont  [1]{#1}%
\providecommand \bibfnamefont [1]{#1}%
\providecommand \citenamefont [1]{#1}%
\providecommand \href@noop [0]{\@secondoftwo}%
\providecommand \href [0]{\begingroup \@sanitize@url \@href}%
\providecommand \@href[1]{\@@startlink{#1}\@@href}%
\providecommand \@@href[1]{\endgroup#1\@@endlink}%
\providecommand \@sanitize@url [0]{\catcode `\\12\catcode `\$12\catcode
  `\&12\catcode `\#12\catcode `\^12\catcode `\_12\catcode `\%12\relax}%
\providecommand \@@startlink[1]{}%
\providecommand \@@endlink[0]{}%
\providecommand \url  [0]{\begingroup\@sanitize@url \@url }%
\providecommand \@url [1]{\endgroup\@href {#1}{\urlprefix }}%
\providecommand \urlprefix  [0]{URL }%
\providecommand \Eprint [0]{\href }%
\providecommand \doibase [0]{https://doi.org/}%
\providecommand \selectlanguage [0]{\@gobble}%
\providecommand \bibinfo  [0]{\@secondoftwo}%
\providecommand \bibfield  [0]{\@secondoftwo}%
\providecommand \translation [1]{[#1]}%
\providecommand \BibitemOpen [0]{}%
\providecommand \bibitemStop [0]{}%
\providecommand \bibitemNoStop [0]{.\EOS\space}%
\providecommand \EOS [0]{\spacefactor3000\relax}%
\providecommand \BibitemShut  [1]{\csname bibitem#1\endcsname}%
\let\auto@bib@innerbib\@empty
\bibitem [{\citenamefont {Christensen}\ and\ \citenamefont {von
  Lilienfeld}(2020)}]{Christensen2020/10.1088/2632-2153/abba6f}%
  \BibitemOpen
  \bibfield  {author} {\bibinfo {author} {\bibfnamefont {A.~S.}\ \bibnamefont
  {Christensen}}\ and\ \bibinfo {author} {\bibfnamefont {O.~A.}\ \bibnamefont
  {von Lilienfeld}},\ }\href {https://doi.org/10.1088/2632-2153/abba6f}
  {\bibfield  {journal} {\bibinfo  {journal} {Mach. Learn.: Sci. Technol.}\
  }\textbf {\bibinfo {volume} {1}},\ \bibinfo {pages} {045018} (\bibinfo {year}
  {2020})}\BibitemShut {NoStop}%
\bibitem [{\citenamefont {Pearson}(1895)}]{Pearson1895/10.1098/rspl.1895.0041}%
  \BibitemOpen
  \bibfield  {author} {\bibinfo {author} {\bibfnamefont {K.}~\bibnamefont
  {Pearson}},\ }\href {https://doi.org/10.1098/rspl.1895.0041} {\bibfield
  {journal} {\bibinfo  {journal} {Proc. R. Soc. Lond.}\ }\textbf {\bibinfo
  {volume} {58}},\ \bibinfo {pages} {240} (\bibinfo {year} {1895})}\BibitemShut
  {NoStop}%
\bibitem [{\citenamefont {Kellner}\ and\ \citenamefont
  {Ceriotti}(2024)}]{Kellner2024/10.1088/2632-2153/ad594a}%
  \BibitemOpen
  \bibfield  {author} {\bibinfo {author} {\bibfnamefont {M.}~\bibnamefont
  {Kellner}}\ and\ \bibinfo {author} {\bibfnamefont {M.}~\bibnamefont
  {Ceriotti}},\ }\href {https://doi.org/10.1088/2632-2153/ad594a} {\bibfield
  {journal} {\bibinfo  {journal} {Mach. Learn.: Sci. Technol.}\ }\textbf
  {\bibinfo {volume} {5}},\ \bibinfo {pages} {035006} (\bibinfo {year}
  {2024})}\BibitemShut {NoStop}%
\bibitem [{\citenamefont {Kovács}\ \emph {et~al.}(2021)\citenamefont
  {Kovács}, \citenamefont {van~der Oord}, \citenamefont {Kucera},
  \citenamefont {Allen}, \citenamefont {Cole}, \citenamefont {Ortner},\ and\
  \citenamefont {Csányi}}]{Kovacs2021/10.1021/acs.jctc.1c00647}%
  \BibitemOpen
  \bibfield  {author} {\bibinfo {author} {\bibfnamefont {D.~P.}\ \bibnamefont
  {Kovács}}, \bibinfo {author} {\bibfnamefont {C.}~\bibnamefont {van~der
  Oord}}, \bibinfo {author} {\bibfnamefont {J.}~\bibnamefont {Kucera}},
  \bibinfo {author} {\bibfnamefont {A.~E.~A.}\ \bibnamefont {Allen}}, \bibinfo
  {author} {\bibfnamefont {D.~J.}\ \bibnamefont {Cole}}, \bibinfo {author}
  {\bibfnamefont {C.}~\bibnamefont {Ortner}},\ and\ \bibinfo {author}
  {\bibfnamefont {G.}~\bibnamefont {Csányi}},\ }\href
  {https://doi.org/10.1021/acs.jctc.1c00647} {\bibfield  {journal} {\bibinfo
  {journal} {J. Chem. Theory Comput.}\ }\textbf {\bibinfo {volume} {17}},\
  \bibinfo {pages} {7696} (\bibinfo {year} {2021})}\BibitemShut {NoStop}%
\bibitem [{\citenamefont {Deng}\ \emph {et~al.}(2023)\citenamefont {Deng},
  \citenamefont {Zhong}, \citenamefont {Jun}, \citenamefont {Riebesell},
  \citenamefont {Han}, \citenamefont {Bartel},\ and\ \citenamefont
  {Ceder}}]{Deng2023/10.1038/s42256-023-00716-3}%
  \BibitemOpen
  \bibfield  {author} {\bibinfo {author} {\bibfnamefont {B.}~\bibnamefont
  {Deng}}, \bibinfo {author} {\bibfnamefont {P.}~\bibnamefont {Zhong}},
  \bibinfo {author} {\bibfnamefont {K.~J.}\ \bibnamefont {Jun}}, \bibinfo
  {author} {\bibfnamefont {J.}~\bibnamefont {Riebesell}}, \bibinfo {author}
  {\bibfnamefont {K.}~\bibnamefont {Han}}, \bibinfo {author} {\bibfnamefont
  {C.~J.}\ \bibnamefont {Bartel}},\ and\ \bibinfo {author} {\bibfnamefont
  {G.}~\bibnamefont {Ceder}},\ }\href
  {https://doi.org/10.1038/s42256-023-00716-3} {\bibfield  {journal} {\bibinfo
  {journal} {Nat. Mach. Intell.}\ }\textbf {\bibinfo {volume} {5}},\ \bibinfo
  {pages} {1031} (\bibinfo {year} {2023})}\BibitemShut {NoStop}%
\bibitem [{\citenamefont {Batatia}\ \emph {et~al.}(2023)\citenamefont
  {Batatia}, \citenamefont {Benner}, \citenamefont {Chiang}, \citenamefont
  {Elena}, \citenamefont {Kovács}, \citenamefont {Riebesell}, \citenamefont
  {Advincula}, \citenamefont {Asta}, \citenamefont {Avaylon}, \citenamefont
  {Baldwin}, \citenamefont {Berger}, \citenamefont {Bernstein}, \citenamefont
  {Bhowmik}, \citenamefont {Blau}, \citenamefont {Cărare}, \citenamefont
  {Darby}, \citenamefont {De}, \citenamefont {Pia}, \citenamefont {Deringer},
  \citenamefont {Elijošius}, \citenamefont {El-Machachi}, \citenamefont
  {Falcioni}, \citenamefont {Fako}, \citenamefont {Ferrari}, \citenamefont
  {Genreith-Schriever}, \citenamefont {George}, \citenamefont {Goodall},
  \citenamefont {Grey}, \citenamefont {Grigorev}, \citenamefont {Han},
  \citenamefont {Handley}, \citenamefont {Heenen}, \citenamefont {Hermansson},
  \citenamefont {Holm}, \citenamefont {Jaafar}, \citenamefont {Hofmann},
  \citenamefont {Jakob}, \citenamefont {Jung}, \citenamefont {Kapil},
  \citenamefont {Kaplan}, \citenamefont {Karimitari}, \citenamefont {Kermode},
  \citenamefont {Kroupa}, \citenamefont {Kullgren}, \citenamefont {Kuner},
  \citenamefont {Kuryla}, \citenamefont {Liepuoniute}, \citenamefont {Margraf},
  \citenamefont {Magdău}, \citenamefont {Michaelides}, \citenamefont {Moore},
  \citenamefont {Naik}, \citenamefont {Niblett}, \citenamefont {Norwood},
  \citenamefont {O'Neill}, \citenamefont {Ortner}, \citenamefont {Persson},
  \citenamefont {Reuter}, \citenamefont {Rosen}, \citenamefont {Schaaf},
  \citenamefont {Schran}, \citenamefont {Shi}, \citenamefont {Sivonxay},
  \citenamefont {Stenczel}, \citenamefont {Svahn}, \citenamefont {Sutton},
  \citenamefont {Swinburne}, \citenamefont {Tilly}, \citenamefont {van~der
  Oord}, \citenamefont {Varga-Umbrich}, \citenamefont {Vegge}, \citenamefont
  {Vondrák}, \citenamefont {Wang}, \citenamefont {Witt}, \citenamefont
  {Zills},\ and\ \citenamefont
  {Csányi}}]{Batatia2023/10.48550/arXiv.2401.00096}%
  \BibitemOpen
  \bibfield  {author} {\bibinfo {author} {\bibfnamefont {I.}~\bibnamefont
  {Batatia}}, \bibinfo {author} {\bibfnamefont {P.}~\bibnamefont {Benner}},
  \bibinfo {author} {\bibfnamefont {Y.}~\bibnamefont {Chiang}}, \bibinfo
  {author} {\bibfnamefont {A.~M.}\ \bibnamefont {Elena}}, \bibinfo {author}
  {\bibfnamefont {D.~P.}\ \bibnamefont {Kovács}}, \bibinfo {author}
  {\bibfnamefont {J.}~\bibnamefont {Riebesell}}, \bibinfo {author}
  {\bibfnamefont {X.~R.}\ \bibnamefont {Advincula}}, \bibinfo {author}
  {\bibfnamefont {M.}~\bibnamefont {Asta}}, \bibinfo {author} {\bibfnamefont
  {M.}~\bibnamefont {Avaylon}}, \bibinfo {author} {\bibfnamefont {W.~J.}\
  \bibnamefont {Baldwin}}, \bibinfo {author} {\bibfnamefont {F.}~\bibnamefont
  {Berger}}, \bibinfo {author} {\bibfnamefont {N.}~\bibnamefont {Bernstein}},
  \bibinfo {author} {\bibfnamefont {A.}~\bibnamefont {Bhowmik}}, \bibinfo
  {author} {\bibfnamefont {S.~M.}\ \bibnamefont {Blau}}, \bibinfo {author}
  {\bibfnamefont {V.}~\bibnamefont {Cărare}}, \bibinfo {author} {\bibfnamefont
  {J.~P.}\ \bibnamefont {Darby}}, \bibinfo {author} {\bibfnamefont
  {S.}~\bibnamefont {De}}, \bibinfo {author} {\bibfnamefont {F.~D.}\
  \bibnamefont {Pia}}, \bibinfo {author} {\bibfnamefont {V.~L.}\ \bibnamefont
  {Deringer}}, \bibinfo {author} {\bibfnamefont {R.}~\bibnamefont
  {Elijošius}}, \bibinfo {author} {\bibfnamefont {Z.}~\bibnamefont
  {El-Machachi}}, \bibinfo {author} {\bibfnamefont {F.}~\bibnamefont
  {Falcioni}}, \bibinfo {author} {\bibfnamefont {E.}~\bibnamefont {Fako}},
  \bibinfo {author} {\bibfnamefont {A.~C.}\ \bibnamefont {Ferrari}}, \bibinfo
  {author} {\bibfnamefont {A.}~\bibnamefont {Genreith-Schriever}}, \bibinfo
  {author} {\bibfnamefont {J.}~\bibnamefont {George}}, \bibinfo {author}
  {\bibfnamefont {R.~E.~A.}\ \bibnamefont {Goodall}}, \bibinfo {author}
  {\bibfnamefont {C.~P.}\ \bibnamefont {Grey}}, \bibinfo {author}
  {\bibfnamefont {P.}~\bibnamefont {Grigorev}}, \bibinfo {author}
  {\bibfnamefont {S.}~\bibnamefont {Han}}, \bibinfo {author} {\bibfnamefont
  {W.}~\bibnamefont {Handley}}, \bibinfo {author} {\bibfnamefont {H.~H.}\
  \bibnamefont {Heenen}}, \bibinfo {author} {\bibfnamefont {K.}~\bibnamefont
  {Hermansson}}, \bibinfo {author} {\bibfnamefont {C.}~\bibnamefont {Holm}},
  \bibinfo {author} {\bibfnamefont {J.}~\bibnamefont {Jaafar}}, \bibinfo
  {author} {\bibfnamefont {S.}~\bibnamefont {Hofmann}}, \bibinfo {author}
  {\bibfnamefont {K.~S.}\ \bibnamefont {Jakob}}, \bibinfo {author}
  {\bibfnamefont {H.}~\bibnamefont {Jung}}, \bibinfo {author} {\bibfnamefont
  {V.}~\bibnamefont {Kapil}}, \bibinfo {author} {\bibfnamefont {A.~D.}\
  \bibnamefont {Kaplan}}, \bibinfo {author} {\bibfnamefont {N.}~\bibnamefont
  {Karimitari}}, \bibinfo {author} {\bibfnamefont {J.~R.}\ \bibnamefont
  {Kermode}}, \bibinfo {author} {\bibfnamefont {N.}~\bibnamefont {Kroupa}},
  \bibinfo {author} {\bibfnamefont {J.}~\bibnamefont {Kullgren}}, \bibinfo
  {author} {\bibfnamefont {M.~C.}\ \bibnamefont {Kuner}}, \bibinfo {author}
  {\bibfnamefont {D.}~\bibnamefont {Kuryla}}, \bibinfo {author} {\bibfnamefont
  {G.}~\bibnamefont {Liepuoniute}}, \bibinfo {author} {\bibfnamefont {J.~T.}\
  \bibnamefont {Margraf}}, \bibinfo {author} {\bibfnamefont {I.-B.}\
  \bibnamefont {Magdău}}, \bibinfo {author} {\bibfnamefont {A.}~\bibnamefont
  {Michaelides}}, \bibinfo {author} {\bibfnamefont {J.~H.}\ \bibnamefont
  {Moore}}, \bibinfo {author} {\bibfnamefont {A.~A.}\ \bibnamefont {Naik}},
  \bibinfo {author} {\bibfnamefont {S.~P.}\ \bibnamefont {Niblett}}, \bibinfo
  {author} {\bibfnamefont {S.~W.}\ \bibnamefont {Norwood}}, \bibinfo {author}
  {\bibfnamefont {N.}~\bibnamefont {O'Neill}}, \bibinfo {author} {\bibfnamefont
  {C.}~\bibnamefont {Ortner}}, \bibinfo {author} {\bibfnamefont {K.~A.}\
  \bibnamefont {Persson}}, \bibinfo {author} {\bibfnamefont {K.}~\bibnamefont
  {Reuter}}, \bibinfo {author} {\bibfnamefont {A.~S.}\ \bibnamefont {Rosen}},
  \bibinfo {author} {\bibfnamefont {L.~L.}\ \bibnamefont {Schaaf}}, \bibinfo
  {author} {\bibfnamefont {C.}~\bibnamefont {Schran}}, \bibinfo {author}
  {\bibfnamefont {B.~X.}\ \bibnamefont {Shi}}, \bibinfo {author} {\bibfnamefont
  {E.}~\bibnamefont {Sivonxay}}, \bibinfo {author} {\bibfnamefont {T.~K.}\
  \bibnamefont {Stenczel}}, \bibinfo {author} {\bibfnamefont {V.}~\bibnamefont
  {Svahn}}, \bibinfo {author} {\bibfnamefont {C.}~\bibnamefont {Sutton}},
  \bibinfo {author} {\bibfnamefont {T.~D.}\ \bibnamefont {Swinburne}}, \bibinfo
  {author} {\bibfnamefont {J.}~\bibnamefont {Tilly}}, \bibinfo {author}
  {\bibfnamefont {C.}~\bibnamefont {van~der Oord}}, \bibinfo {author}
  {\bibfnamefont {E.}~\bibnamefont {Varga-Umbrich}}, \bibinfo {author}
  {\bibfnamefont {T.}~\bibnamefont {Vegge}}, \bibinfo {author} {\bibfnamefont
  {M.}~\bibnamefont {Vondrák}}, \bibinfo {author} {\bibfnamefont
  {Y.}~\bibnamefont {Wang}}, \bibinfo {author} {\bibfnamefont {W.~C.}\
  \bibnamefont {Witt}}, \bibinfo {author} {\bibfnamefont {F.}~\bibnamefont
  {Zills}},\ and\ \bibinfo {author} {\bibfnamefont {G.}~\bibnamefont
  {Csányi}},\ }\href {https://doi.org/10.48550/arXiv.2401.00096} {\bibfield
  {journal} {\bibinfo  {journal} {arxiv preprint}\ } (\bibinfo {year} {2023}),\
  10.48550/arXiv.2401.00096},\ \Eprint {https://arxiv.org/abs/2401.00096}
  {2401.00096} \BibitemShut {NoStop}%
\bibitem [{\citenamefont {Schran}, \citenamefont {Brezina},\ and\ \citenamefont
  {Marsalek}(2020)}]{Schran2020/10.1063/5.0016004}%
  \BibitemOpen
  \bibfield  {author} {\bibinfo {author} {\bibfnamefont {C.}~\bibnamefont
  {Schran}}, \bibinfo {author} {\bibfnamefont {K.}~\bibnamefont {Brezina}},\
  and\ \bibinfo {author} {\bibfnamefont {O.}~\bibnamefont {Marsalek}},\ }\href
  {https://doi.org/10.1063/5.0016004} {\bibfield  {journal} {\bibinfo
  {journal} {J. Chem. Phys.}\ }\textbf {\bibinfo {volume} {153}} (\bibinfo
  {year} {2020}),\ 10.1063/5.0016004}\BibitemShut {NoStop}%
\bibitem [{\citenamefont {Barroso-Luque}\ \emph {et~al.}(2024)\citenamefont
  {Barroso-Luque}, \citenamefont {Muhammed}, \citenamefont {Fu}, \citenamefont
  {Brandon}, \citenamefont {Gao}, \citenamefont {Rizvi}, \citenamefont
  {Zitnick},\ and\ \citenamefont
  {Ulissi}}]{Barroso-Luque/10.48550/arXiv.2410.12771}%
  \BibitemOpen
  \bibfield  {author} {\bibinfo {author} {\bibfnamefont {L.}~\bibnamefont
  {Barroso-Luque}}, \bibinfo {author} {\bibfnamefont {S.}~\bibnamefont
  {Muhammed}}, \bibinfo {author} {\bibfnamefont {X.}~\bibnamefont {Fu}},
  \bibinfo {author} {\bibfnamefont {D.~M.~W.}\ \bibnamefont {Brandon}},
  \bibinfo {author} {\bibfnamefont {M.}~\bibnamefont {Gao}}, \bibinfo {author}
  {\bibfnamefont {A.}~\bibnamefont {Rizvi}}, \bibinfo {author} {\bibfnamefont
  {C.~L.}\ \bibnamefont {Zitnick}},\ and\ \bibinfo {author} {\bibfnamefont
  {Z.~W.}\ \bibnamefont {Ulissi}},\ }\href
  {https://doi.org/10.48550/arXiv.2410.12771} {\bibfield  {journal} {\bibinfo
  {journal} {arXiv preprint}\ } (\bibinfo {year} {2024}),\
  10.48550/arXiv.2410.12771},\ \Eprint {https://arxiv.org/abs/2410.12771}
  {arXiv:2410.12771 [cond-mat.mtrl-sci]} \BibitemShut {NoStop}%
\bibitem [{\citenamefont {Kaur}\ \emph {et~al.}(2025)\citenamefont {Kaur},
  \citenamefont {Pia}, \citenamefont {Batatia}, \citenamefont {Advincula},
  \citenamefont {Shi}, \citenamefont {Lan}, \citenamefont {Csányi},
  \citenamefont {Michaelides},\ and\ \citenamefont
  {Kapil}}]{Kaur2025/10.1039/D4FD00107A}%
  \BibitemOpen
  \bibfield  {author} {\bibinfo {author} {\bibfnamefont {H.}~\bibnamefont
  {Kaur}}, \bibinfo {author} {\bibfnamefont {F.~D.}\ \bibnamefont {Pia}},
  \bibinfo {author} {\bibfnamefont {I.}~\bibnamefont {Batatia}}, \bibinfo
  {author} {\bibfnamefont {X.~R.}\ \bibnamefont {Advincula}}, \bibinfo {author}
  {\bibfnamefont {B.~X.}\ \bibnamefont {Shi}}, \bibinfo {author} {\bibfnamefont
  {J.}~\bibnamefont {Lan}}, \bibinfo {author} {\bibfnamefont {G.}~\bibnamefont
  {Csányi}}, \bibinfo {author} {\bibfnamefont {A.}~\bibnamefont
  {Michaelides}},\ and\ \bibinfo {author} {\bibfnamefont {V.}~\bibnamefont
  {Kapil}},\ }\href {https://doi.org/10.1039/D4FD00107A} {\bibfield  {journal}
  {\bibinfo  {journal} {Faraday Discuss.}\ }\textbf {\bibinfo {volume} {256}},\
  \bibinfo {pages} {120} (\bibinfo {year} {2025})}\BibitemShut {NoStop}%
\end{thebibliography}%
\end{document}